\documentclass[12pt,leqno,fleqn]{amsart}
\usepackage{latexsym}
\usepackage{amssymb}
\usepackage{amsxtra}
\usepackage{amsmath}
\usepackage{amsfonts}
\usepackage[dvips]{graphicx}
\usepackage{amsmath}
\usepackage{amssymb,amsmath,amsthm,amscd,mathrsfs,amsbsy,amsfonts}
\usepackage{pstricks,pst-node}
\usepackage{amsbsy,young,colortbl,cite}

\usepackage{graphicx}
\usepackage[metapost]{mfpic}
\usepackage{mflogo}

\usepackage{epstopdf}
\usepackage{graphics}
\def\@makecaption#1#2{\vskip\abovecaptionskip
  \sbox\@tempboxa{\small #1: #2}%
  \ifdim \wd\@tempboxa >\hsize \small #1: #2\par
  \else \global \@minipagefalse \hb@xt@\hsize{\hfil\box\@tempboxa\hfil}\fi
  \vskip\belowcaptionskip}
\newcommand{\cleqn}{\setcounter{equation}{0}}
\newcommand{\clth}{\setcounter{theorem}{0}}
\newcommand {\sectionnew}[1]{\section{#1}\cleqn\clth}
\newtheorem{theorem}{Theorem}[section]
\newtheorem{lemma}[theorem]{Lemma}
\newtheorem{proposition}[theorem]{Proposition}

\newtheorem{definition}[theorem]{Definition}

\newtheorem{Example}[theorem]{Example}

\theoremstyle{remark}

\renewcommand{\P}{\mathcal{P}}
\newcommand{\W}{\mathcal{W}}
\renewcommand{\L}{\mathcal{L}}

\newcommand{\Z}{\mathbb{Z}}
\newcommand{\N}{\mathbb{N}}
\renewcommand{\H}{\mathcal{H}}

\def\res{\mathop{\rm Res}\nolimits}
\def\({\left(}
\def\){\right)}
\def\[{\left[}
\def\]{\right]}
\def\d{\partial}
\def\ep{\epsilon}
\def\La{\Lambda}
\def\la{\lambda}

\begin{document}

\title{Solutions of  bigraded Toda Hierarchy}
\author{Chuanzhong Li}
\dedicatory {Department of Mathematics, USTC, Hefei, 230026, Anhui, P. R. China\\
 Department of
Mathematics,  NBU, Ningbo, 315211, Zhejiang, P. R. China}

\thanks{Correspondence: czli@mail.ustc.edu.cn}
\texttt{}

\date{}
\begin{abstract}
The $(N,M)$-bigraded Toda hierarchy is an extension of the original Toda lattice hierarchy.
The pair of numbers $(N,M)$ represents the band structure of the Lax matrix which has $N$ upper
and $M$ lower diagonals,
and the original one is referred to as the $(1,1)$-bigraded Toda hierarchy.
 Because of this band structure, one can introduce
$M+N-1$ commuting flows which give a parametrization of a small phase space for a topological field theory.

In this paper, we first show that there exists a natural symmetry between the $(N,M)$- and
$(M,N)$-bigraded Toda hierarchies.  We then derive the Hirota bilinear form for those commuting flows,
which consists of two-dimensional Toda hierarchy, the discrete KP hierarchy and its B\"acklund transformations.
We also discuss the solution structure of the $(N,M)$-bigraded Toda equation in terms of the moment
matrix defined via the wave operators associated with the Lax operator, and construct
some of the explicit solutions.  In particular, we give the rational solutions which are expressed by the products of
the Schur polynomials corresponding to non-rectangular Young diagrams.
\end{abstract}
\maketitle
\tableofcontents
\allowdisplaybreaks
 \setcounter{section}{0}

\section{Introduction}

The $(N,M)$-bigraded Toda hierarchy, denoted by $(N,M)$-BTH, is an integrable system (see e.g.
\cite{C, our}), and its Lax operator is given by
\begin{equation*}
\L:=\Lambda^{N}+u_{N-1}\Lambda^{N-1}+\cdots+u_0+\cdots + u_{-M}
\Lambda^{-M}.
\end{equation*}
where $N,M \geq1$, $\Lambda$ is a shift operator which can be expressed as an infinite matrix in the form,
$\Lambda=(E_{i,i+1})_{i\in\mathbb{Z}}$.  In terms of an infinite size matrix, the Lax operator $\mathcal{L}$ has
the band structure with $N$ upper and $M$ lower nonzero diagonals.
The $(N,M)$-BTH is then defined by
\begin{equation}\label{BTHLax'}
\frac{\partial \L}{\partial t_{\gamma, n}} =\left\{
\begin{array}{lll}
[ (\L^{n+1-\frac{\alpha-1}{N}})_+, \L ], &\quad {\rm if} \quad\gamma=\alpha= 1,2,\ldots,N,\\[1.5ex]
[ -( \L^{n+1+\frac{\beta}{M}})_-, \L ], &\quad{\rm if}\quad \gamma=\beta=   -M+1,\ldots,-1,0,
\end{array}\right.
\end{equation}
Here we call the $N+M-1$ numbers of the flows for $n=0$ the {\it primaries} of the BTH
which describe the small phase space of a topological field theory (TFT), and the flows with $n>0$
correspond to the gravitational descendants in this TFT.

In the case of $N=M=1$, the $(1,1)$-BTH is the original Toda lattice hierarchy, and the primary flow
is just the Toda lattice equation \cite{Toda,Todabook}.  The $(N,M)$-BTH with $N>1$ or $M>1$ is then considered as an extension
of the original Toda lattice hierarchy. One should also note that the case with infinite $N$ and $M$ corresponds
to the two-dimensional Toda hierarchy where we have two independent Lax operators (one defined
near infinity and the other defined near zero in the spectral space, more precisely, one considers two cases
with $(1,\infty)$ and $(\infty,1)$).
Then the $(N,M)$-BTH can be naturally considered as a reduction of the two-dimensional Toda hierarchy
by imposing an algebraic relation to those two Lax operators (see \cite{Takasaki, UT, ourBlock}).

In \cite{our}, we showed the integrability of an extended version of $(N,M)$-BTH by writing the hierarchy as a bilinear identity, and introduced the $\tau$-functions.  Here the extension implies that the hierarchy has additional
logarithmic flows, and this version is called the extended BTH (see \cite{TH, C}).  In this paper, we are interested in constructing several explicit solutions of the BTH, and as a first step, we only consider the non-extended version of the BTH
based on our previous study \cite{our}.

The paper is organized as follows.  In section \ref{sec:Toda}, we give a brief summary of the original Toda lattice
hierarchy whose Lax operator is given by a tri-diagonal matrix.  We also discuss briefly $(2,1)$-BTH
as an extension of the Toda hierarchy and describe the $t_{2,0}$-flow defined by the square root of
the Lax operator. In particular, we mention that there exists nonlocal terms in the equation,
and make a remark that the flow also appears in the recent paper \cite{Blaszak-Szum}.
In section \ref{sec:BTH}, we give the explicit form of $(N,M)$-BTH and the $\tau$-functions. This section is
a brief summary of our previous paper \cite{our} without the logarithmic flows.  Here we express the coefficient
functions in the Lax operator in terms of the $\tau$-functions in the similar manner discussed in \cite{DiscreteKP, adler92}.
We also discuss some details of the $(2,2)$-BTH as an example.
In section \ref{sec:Equivalent}, we show the equivalence between the $(N,M)$- and $(M,N)$-BTHs by
using the Hirota bilinear equations found in \cite{our} and gauge transformation in \cite{BlaszakMarciniakJMP}.
 To be illustrative, we also give some simplest concrete examples of
equivalence between flows of $(2,1)$-BTH and
$(1,2)$-BTH. This equivalence is explicitly shown in
the examples given in section \ref{sec:Hirota}.
In section \ref{sec:Hirota}, we derive the Hirota bilinear equations for the primaries of the $(N,M)$-BTH.
Then we construct the $\tau$-functions in terms of the moment matrix defined naturally via the wave operators
introduced in section \ref{sec:BTH} (also see \cite{our}).
In section \ref{sec:Rational}, we construct rational solutions based on the $\tau$-function formulas derived in
the previous section. In particular, those rational solutions are given by the products of two Schur polynomials
depending on two different sets of flow parameters $t_{\alpha,n}$ and $t_{\beta,n}$ in (\ref{BTHLax'}).
Contrary to the case of the original Toda hierarchy where the rational solutions are given by
the Schur polynomials of rectangular Young diagrams, the rational solutions of the BTH are parametrized by
non-rectangular Young diagrams.
Finally, in section \ref{sec:Conclusion}, we summarize the results and give some discussions.

\section{Tridigonal Toda lattice hierarchy and generalization}\label{sec:Toda}
Here we briefly explain the BTH as an extension of the original Toda equation, and
present some connection to the recent study in \cite{Blaszak-Szum}.
The main point is to explain the structure of an additional symmetry generated by a fractional power of
the Lax matrix.
The Toda lattice equation is written in the form,
\begin{equation}\label{toda}\left\{
\begin{array}{lllll}
\displaystyle{\frac{\partial a_{n+1}}{\partial t_1}}&=a_{n+1}(b_{n+1}-b_{n}),\\[2.0ex]
\displaystyle{\frac{\partial b_n}{\partial t_1}}&=a_{n+1}-a_{n}\,.
\end{array}\right. \qquad n=1,2,\ldots.
\end{equation}
Eq.\eqref{toda} has the Lax representation with a  tridiagonal semi-infinite matrix $L$ by
\begin{equation}
\frac{\partial L}{\partial t_1}=[B_1, L],\qquad B_1=[L]_{\ge 0}\,,
\end{equation}
where $[L]_{\ge 0}$ is the upper triangular part of the matrix $L$ given by
\begin{equation}
L=\begin{pmatrix}
b_1 &  1 &  0  & 0 & \cdots  \\
a_2 & b_2 & 1 & 0 & \cdots \\
0  & a_3 & b_3 & 1 & \cdots\\
0 &  0 & a_4 & b_4 & \cdots \\
\vdots&\vdots &\vdots&\vdots& \ddots
\end{pmatrix}\,.
\end{equation}
If we consider the bounded Toda lattice equation, the Lax matrix will have finite size.

For semi-infinite Toda equation there exists a sequence  of $\tau$-functions $\{\tau_n:n\ge 0\}$  with $\tau_0=1$ defined by the $a_n, b_n$ by the formulas
\begin{equation}
a_n=\frac{\tau_{n}\tau_{n-2}}{\tau_{n-1}^2},\qquad b_n=\frac{\partial}{\partial t_1}\log \left(\frac{\tau_n}{\tau_{n-1}}\right)\,.
\end{equation}
Then we can write the Toda lattice equation in the Hirota bilinear form,
\begin{equation}\label{Toda-Hirota}
D_1^2\tau_n\cdot\tau_n=2\tau_{n+1}\tau_{n-1},
\end{equation}
where $D_1$ is the usual Hirota derivative.
For the $k$-th  flow-parameter $t_k$  of the Toda lattice hierarchy, $D_k$ is defined by
\begin{equation}
D_kf\cdot g:=\left(\frac{\partial}{\partial t_k}-\frac{\partial }{\partial t_k'}\right) f(t_k)g(t_k')\Big|_{t_k=t_k'}\,.
\end{equation}

The hierarchy of the Toda lattice is defined by
\begin{equation}
\frac{\partial L}{\partial t_k}=[B_k, L],\qquad  B_k=[L^k]_{\ge 0},\qquad k=1,2,3,\ldots\,.
\end{equation}

The $\tau$-functions of the Toda lattice hierarchy obey the following
equations
\begin{equation}
[D_{k}-P_{k}({\hat D})]\tau_{n+1}\cdot \tau_{n}=0, \quad k=2,3,4,
... ,
\end{equation}
where Schur polynomial $P_{k}({\hat D})$ is defined by
\begin{equation}\label{shurfunction}
e^{\sum^{\infty}_{k=1}\frac{1}{k}D_kz^k}=\sum^{\infty}_{k=0}P_{k}({\hat
D})z^{k}, \quad {\hat D}=(D_1, \frac{1}{2}D_2, \frac{1}{3}D_3,
\frac{1}{4}D_4, ... ).
\end{equation}
When $k=2$, Hirota equation becomes
\begin{equation}\label{NLS}
(D_2-D^2_1)\tau_{n+1}\cdot\tau_{n}=0.
\end{equation}
Eq.\eqref{NLS} together with eq.\eqref{Toda-Hirota}
will give  the nonlinear Schrodinger equation
 which can be seen as the second member of the Toda lattice hierarchy \cite{Kodama-Pierce}.

A natural question will be how about generalized band structure of the Toda Lattice hierarchy which is just $(1,1)$ tridiagonal band matrix.

For example, for $(2,1)$ Heissenberg band structure of Lax matrix
\begin{equation}
\tilde L=\begin{pmatrix}
b_1 &  c_1 &  1  & 0 & \cdots  \\
a_2 & b_2 & c_2 & 1 & \cdots \\
0  & a_3 & b_3 & c_3 & \cdots\\
0 &  0 & a_4 & b_4 & \cdots \\
\vdots&\vdots &\vdots&\vdots& \ddots
\end{pmatrix},
\end{equation}
the equations for Toda flow, i.e.
\begin{eqnarray}
 \partial_{t_{1,0}} \tilde L= [\tilde L_{\geq 0}, \tilde L],
\end{eqnarray} will lead to the following Blaszak-Marciniak lattice equation \cite{BlaszakMarciniakJMP}
\begin{eqnarray}\label{21t_{1,0}}
\begin{cases}
\partial_{t_{1,0}} c_n&= a_{n+2}-a_n\\
 \partial_{t_{1,0}} b_n&= c_na_{n+1}-a_nc_{n-1}\\
  \partial_{t_{1,0}} a_n&=a_n( b_n-b_{n-1}).
   \end{cases}
\end{eqnarray}
Eq.\eqref{21t_{1,0}} is also equivalent to the
Bogoyavlensky--Narita lattice given in \cite{SvininJPA}. In \cite{BlaszakMarciniakJMP},
Blaszak and Marciniak  considered the local flows which correspond to the integer powers of Lax operators. In this paper, we construct nonlocal flows using fractional power of Lax operator.
Because of the $(2,1)$-band structure, one can define the square root  of Lax matrix which will be shown in detail in the next section. Using the operator $\tilde L^{\frac12}$, we give a new flow
\begin{eqnarray}
\partial_{t_{2,0}} \tilde L= [\tilde L^{\frac12}_{\geq 0}, \tilde L],
\end{eqnarray}
which further leads to
\begin{eqnarray}\label{2,0}
\begin{cases}
\partial_{t_{2,0}} c_n&= b_{n+1}-b_n+c_n(1-\La)(1+\La)^{-1}c_n\\
 \partial_{t_{2,0}} b_n&= a_{n+1}-a_n\\
 \partial_{t_{2,0}} a_n&= a_n(1-\La^{-1})(1+\La)^{-1}c_n.
 \end{cases}
\end{eqnarray}

After denoting $\H$ as $\frac{1+\La}{\La-1}$, eq.\eqref{2,0} can be rewritten as
\begin{eqnarray}\label{2,0flow}
\begin{cases}
\partial_{t_{2,0}} c_n&= b_{n+1}-b_n+c_n\H^{-1}c_n\\
 \partial_{t_{2,0}} b_n&= a_{n+1}-a_n\\
 \partial_{t_{2,0}} a_n&= a_n\H^{-1}c_n.
 \end{cases}
\end{eqnarray}
After transformation
\begin{eqnarray}
 c_n=\bar c_{n+1}+\bar c_n,
\end{eqnarray}
eq.\eqref{2,0} becomes
\begin{eqnarray}\label{2,0'}
\begin{cases}
\partial_{t_{2,0}} \bar c_{n+1}+\partial_{t_{2,0}} \bar c_n&= b_{n+1}- b_n+\bar c_n^2-\bar c_{n+1}^2\\
 \partial_{t_{2,0}} b_n&=  a_{n+1}- a_n\\
 \partial_{t_{2,0}}  a_n&=  a_n(\bar c_n-\bar c_{n-1}),
 \end{cases}
\end{eqnarray}
which is just eq.$(40)$ in \cite{SvininJPA} and also related to the system (10)-(12) proposed in \cite{YTWuXBHuJPA}. In \cite{SvininJPA},  Svinin gives general constructions of such flows.  Studying the relation between two ways of constructing nonlocal flows, i.e. the way in \cite{SvininJPA} and the way in this paper might be an interesting
problem.

Eq. \eqref{21t_{1,0}} and eq. \eqref{2,0flow} are in fact $t_1$ flow and $t_2$ flow of the case $n=3, \alpha=-1$  without center extension in \cite{Blaszak-Szum}, i.e. solutions do not depend on $y$ variable. In this paper,  we directly introduce a new fractional Lax matrix instead of using Casimir construction
used in \cite{Blaszak-Szum} for constructing the Lax equations.
We also generalize these results to $(N,M)$-band matrix. For a finite-sized Lax matrix,
its fractional power may not be well-defined. This leads to a difficulty to give symmetric flows generated by fraction powers of Lax matrix. But for a bi-infinite band matrix, one can define the fraction powers and further define other additional flows which commute with the original Toda flow. This generalization leads to the BTH which might be also seen as a general reduction of the two-dimensional Toda lattice hierarchy. In the next section, we give the continuous interpolated version of BTH \cite{our} and later
present the matrix version of the BTH in bi-infinite and semi-infinite band matrices.

\section{The bigraded Toda hierarchy (BTH)}\label{sec:BTH}

The Lax operator of the BTH is given by the Laurent polynomial of $\Lambda$ \cite{C}
\begin{equation}\label{LBTH}
\L:=\Lambda^{N}+u_{N-1}\Lambda^{N-1}+\cdots+u_0+\cdots + u_{-M}
\Lambda^{-M},
\end{equation}
where
   $N,M \geq1$, $\Lambda$ represents the shift operator with $\Lambda:=e^{\epsilon\partial_x}$ and $``\epsilon"$ is
called the string coupling constant, i.e. for any function $ f(x)$
\begin{equation*}
\Lambda f(x)=f(x+\epsilon).
\end{equation*}
 The $\L$ can be
written in two different ways by dressing the shift operator
\begin{equation}\label{two dressing}
\L=\P_L\Lambda^N\P_L^{-1} = \P_R \Lambda^{-M}\P_R^{-1},
\end{equation}
where the dressing operators have the form,
 \begin{align}
 \P_L&=1+w_1\Lambda^{-1}+w_2\Lambda^{-2}+\ldots,
\label{dressP}\\[0.5ex]
 \P_R&=\tilde{w_0}+\tilde{w_1}\Lambda+\tilde{w_2}\Lambda^2+ \ldots.
\label{dressQ}
\end{align}
Eq.\eqref{two dressing} are quite important because it gives the reduction condition from the two-dimensional Toda lattice hierarchy.
 The
pair is unique up to multiplying $\P_L$ and $\P_R$ from the right
 by operators in the form  $1+
a_1\Lambda^{-1}+a_2\Lambda^{-2}+...$ and $\tilde{a}_0 +
\tilde{a}_1\Lambda +\tilde{a}_2\Lambda^2+\ldots$ respectively with
coefficients independent of $x$. Given any difference operator $A=
\sum_k A_k \Lambda^k$, the positive and negative projections are
defined by $A_+ = \sum_{k\geq0} A_k \Lambda^k$ and $A_- = \sum_{k<0}
A_k \Lambda^k$.

To write out  explicitly the Lax equations  of BTH,
  fractional powers $\L^{\frac1N}$ and
$\L^{\frac1M}$ are defined by
\begin{equation}
  \notag
  \L^{\frac1N} = \Lambda+ \sum_{k\leq 0} a_k \Lambda^k , \qquad \L^{\frac1M} = \sum_{k \geq -1} b_k
  \Lambda^k,
\end{equation}
with the relations
\begin{equation}
  \notag
  (\L^{\frac1N} )^N = (\L^{\frac1M} )^M = \L.
\end{equation}
Acting on free function, these two fraction powers can be seen as two different locally expansions around zero and infinity respectively.
It was  stressed that $\L^{\frac1N}$ and $\L^{\frac1M}$ are two
different operators even if $N=M(N, M\geq 2)$ in \cite{C} due to two different dressing operators. They can also be
expressed as following
\begin{equation}
\notag
  \L^{\frac1N} = \P _{L}\Lambda\P_{L}^{-1}, \qquad \L^{\frac1M} = \P_{R}\Lambda^{-1} \P_{R}^{ -1}.
\end{equation}
Let  us now define the following operators for the generators of the BTH flows,
\begin{equation}
  B_{\gamma , n} :=\left\{
\begin{array}{llll}
 \displaystyle{ \L^{n+1-\frac{\alpha-1}{N}} }&\qquad {\rm if}\quad \gamma=\alpha=1,2,\ldots,N\\[2.0ex]
 \displaystyle{   \L^{n+1+\frac{\beta}{M}}} &\qquad{\rm if}\quad\gamma=\beta = -M+1,\ldots,-1,0,
\end{array}\right.
\end{equation}
\begin{definition}
  Bigraded Toda hierarchy (BTH) in the Lax representation is given by the set of
infinite number of flows defined by
\begin{equation}\label{BTHLax}
\frac{\partial \L}{\partial t_{\gamma, n}} =\left\{
\begin{array}{lll}
[ (B_{\alpha,n})_+, \L ], &\quad {\rm if} \quad\gamma=\alpha= 1,2,\ldots,N,\\[1.5ex]
[ -(B_{\beta,n})_-, \L ], &\quad{\rm if}\quad \gamma=\beta=   -M+1,\ldots,-1,0.
\end{array}\right.
\end{equation}
\end{definition}

We need to remark that this kind of definition is equivalent to the definition in \cite{C} which is just a scalar transformation about time variables using gamma function.
The original tridiagonal Toda (i.e. Kostant-Toda) hierarchy corresponds to the case with $N=M=1$.

One can show \cite{our} that the BTH in the Lax representation can be written in the equations
of the dressing operators (i.e. the Sato equations):
\begin{theorem}
\label{t1} The operator $\L$ in \eqref{LBTH} is a solution to the BTH \eqref{BTHLax} if and only if there is a
pair of dressing operators $\P_L$ and $\P_R$ which satisfy the Sato  equations,
\begin{equation}
\label{bn1}
\d_{\gamma,n}\P_L  =- (B_{\gamma,n})_-  \P_L, \qquad  \d_{\gamma,n}\P_R = (B_{\gamma ,n})_+\P_R
\end{equation}
for $-M+1\leq\gamma\leq N$ and  $n \geq 0.$
\end{theorem}
The dressing operators satisfying Sato equations will be called wave operators. By wave operators we will give the definition of tau
function for BTH as following.

According to paper \cite{our},
a function $\tau$  depending only on the dynamical variables $t$ and
$\epsilon$ is called the  {\em \bf tau-function of BTH} if it
provides symbols related to wave operators as following,
\begin{eqnarray}\label{pltau}P_L: &=&
1+\frac{w_1}{\lambda}+\frac{w_2}{\lambda^2}+\ldots : =\frac{ \tau
(x, t-[\lambda^{-1}]^N;\epsilon) }
     {\tau (x,t;\epsilon)},\\\label{pl-1tau}
P_L^{-1}:& = &1+\frac{w_1'}{\lambda}+\frac{w_2'}{\lambda^2}+\ldots
: = \frac{\tau (x+\epsilon,
t+[\lambda^{-1}]^N;\epsilon) }
     {\tau (x+\epsilon,t;\epsilon)},\\\label{prtau}
P_R:&= &\tilde w_0 + \tilde w_1\lambda+ \tilde w_2\lambda^{2}
+\ldots : = \frac{ \tau
(x+\epsilon,t+[\lambda]^M;\epsilon)}
     {\tau(x,t;\epsilon)},\\\label{pr-1tau}
     P_R^{-1}:&= &\tilde w_0 '+ \tilde w_1'\lambda+ \tilde
w_2'\lambda^{2} +\ldots : = \frac{
     \tau (x,t-[\lambda]^M; \epsilon)}
     {\tau(x+\epsilon,t;\epsilon)},
     \end{eqnarray}
     where $[\lambda^{-1}]^N$ and $\[\lambda\]^{M}$  are defined by
\begin{equation} \notag
 \[\lambda^{-1}\]^{N}_{\gamma,n} :=
\begin{cases}
  \frac{\lambda^{-N(n+1-\frac{\gamma-1}{N})}}{N(n+1-\frac{\gamma-1}{N})}, &\gamma=N,N-1,\dots 1,\\
 0, &\gamma = 0, -1\dots -(M-1),
  \end{cases}
\end{equation}
\begin{equation} \notag
\[\lambda\]^{M}_{\gamma,n} :=
\begin{cases}
0, &\gamma=N, N-1,\dots 1,\\
\frac{\lambda^{M(n+1+\frac{\beta}{M})}}{M(n+1+\frac{\beta}{M})}
, &\gamma = 0, -1, \dots -(M-1).
  \end{cases}
\end{equation}
  For a given pair of wave
operators the tau-function is unique up to  a non-vanishing function
factor which is independent of $x$ and
$t_{\gamma,n}$ with all $n\geq 0$ and $-M+1\leq\gamma\leq{N}$.\\
Then we get
\begin{eqnarray}\label{pltau}
&P_L: =\displaystyle{\sum_{n=0}^{\infty}\frac{ P_n(-\hat\d_L)\tau
(x,t;\epsilon) }
     {\tau (x,t;\epsilon)}\la^{-n}},\qquad
&P_L^{-1}:= \sum_{n=0}^{\infty}\frac{P_n(\hat\d_L)\tau (x+\epsilon,
t;\epsilon) }
     {\tau (x+\epsilon,t;\epsilon)}\la^{-n},\\\label{prtau}
&P_R:= \displaystyle{\sum_{n=0}^{\infty}\frac{ P_n(\hat\d_R)\tau
(x+\epsilon,t;\epsilon)}
     {\tau(x,t;\epsilon)}\la^{n}}, \qquad
   &  P_R^{-1}:= \sum_{n=0}^{\infty}\frac{P_n(-\hat\d_R)
     \tau (x,t; \epsilon)}
     {\tau(x+\epsilon,t;\epsilon)}\la^{n},
     \end{eqnarray}
where $P_n$ are the elementary Schur polynomial as defined in \eqref{shurfunction}.
Here the operators $\hat\partial_L$ and $\hat\partial_R$ are defined by
\begin{align*}
\hat \d_L&=\left\{\frac{1}{N(n+1-\frac{\alpha-1}{N})}\d_{t_{\alpha,n}}:  1\leq\alpha\leq N\right\}\\[1.0ex]
\hat \d_R&=\left\{\frac{1}{M(n+1+\frac{\beta}{M})}\d_{t_{\beta,n}}: -M+1\leq\beta\leq0\right\}.
\end{align*}

The dressing operators $\P_L$ and $\P_R$ can be expressed by function $\tau(x,t;\epsilon)$:
 \begin{align}\label{pltau}\P_L&=\sum_{n=0}^{\infty}\frac{ P_n(-\hat\d_L)\tau
(x,t;\epsilon) }
     {\tau (x,t;\epsilon)}\La^{-n},\qquad
\P_L^{-1} =  \sum_{n=0}^{\infty}\La^{-n}\frac{P_n(\hat\d_L)\tau (x+\epsilon,
t;\epsilon) }
     {\tau (x+\epsilon,t;\epsilon)},\\\label{prtau}
\P_R&=  \sum_{n=0}^{\infty}\frac{ P_n(\hat\d_R)\tau
(x+\epsilon,t;\epsilon)}
     {\tau(x,t;\epsilon)}\La^{n},\qquad
     \P_R^{-1}
     =  \sum_{n=0}^{\infty}\La^{n}\frac{P_n(-\hat\d_R)
     \tau (x,t; \epsilon)}
     {\tau(x+\epsilon,t;\epsilon)}.
     \end{align}

 One can then find the explicit form of the coefficients $u_i(x,t)$ of the operator $\mathcal{L}$ in terms of the $\tau$-function using eq.\eqref{two dressing} as \cite{DiscreteKP, adler92},
    \begin{align}\label{ui with tau}
 u_i(x,t)&=\frac{P_{N-i}(\hat D_L)\tau (x+(i+1)\ep,
t;\epsilon)\circ \tau
(x,t;\epsilon) }
     {\tau (x,t;\epsilon)\,\tau (x+(i+1)\ep,t;\epsilon)}=\frac{ P_{M+i}(\hat D_R)\tau
(x+\epsilon,t;\epsilon)\circ\tau (x+i\ep,t; \epsilon)}
     {\tau(x,t;\epsilon)\,\tau(x+(i+1)\ep,t;\epsilon)},
 \end{align}
 where $\hat D_L$ and $\hat D_R$ are just the Hirota derivatives corresponding to $\hat \d_L$ and $\hat \d_R$ respectively.

  The BTH can be also written in the matrix form with the identification of the shift operator $\Lambda$ as
the infinite matrix having zero entries except 1's in the upper diagonal elements and all other functions about $x$ as diagonal infinite matrix\cite{UT}. But now we only consider its reduction, i.e. the corresponding semi-infinite matrix form in the following. Then we rewrite
the coefficient $u_i(x,t)$ as $u_{i,j}(t)$ and rewrite $\tau(x+\epsilon, t)$ as $\tau_j(t).$
We can find the corresponding semi-infinite matrix forms $\tilde \P_L,\tilde \P_L^{-1}, \tilde \P_R, \tilde \P_R^{-1}$  corresponding to  $\P_L,\P_L^{-1}, \P_R, \P_R^{-1}$ respectively as following

   \begin{eqnarray}\label{momentL1}
\tilde\P_L
 &=&\left(\begin{array}{ccccc}1 &0&0&0&\dots \\ \frac{P_1(-\hat \d_L)\tau_1}{\tau_1} &1 &0&0&\dots
\\ \frac{P_2(-\hat \d_L)\tau_2}{\tau_2}&\frac{P_1(-\hat \d_L)\tau_2}{\tau_2} &1 &0&\dots
\\ \frac{P_3(-\hat \d_L)\tau_3}{\tau_3}&\frac{P_2(-\hat \d_L)\tau_3}{\tau_3} &\frac{P_1(-\hat \d_L)\tau_3}{\tau_3} &1&\dots \\ \dots &\dots &\dots &\dots&\dots
\end{array}\right),
  \end{eqnarray}

\begin{eqnarray}\label{momentL}
\tilde\P_L^{-1}&=&\ \ \ \left(\begin{array}{ccccc}1 &0&0&0&\dots \\ \frac{P_1(\hat \d_L)\tau_1}{\tau_1} &1 &0&0&\dots
\\ \frac{P_2(\hat \d_L)\tau_1}{\tau_1}&\frac{P_1(\hat \d_L)\tau_2}{\tau_2} &1 &0&\dots
\\ \frac{P_3(\hat \d_L)\tau_1}{\tau_1}&\frac{P_2(\hat \d_L)\tau_2}{\tau_2} &\frac{P_1(\hat \d_L)\tau_2}{\tau_2} &1&\dots\\ \dots &\dots &\dots &\dots&\dots
\end{array}\right),\\ \label{momentR}
\tilde\P_R&=&\left(\begin{array}{ccccc}\frac{\tau_1}{\tau_0} &\frac{P_1(\hat \d_R)\tau_1}{\tau_0} &\frac{P_2(\hat \d_R)\tau_1}{\tau_0} &\frac{P_3(\hat \d_R)\tau_1}{\tau_0} &\dots \\
0 &\frac{\tau_2}{\tau_1}  &\frac{P_1(\hat \d_R)\tau_2}{\tau_1}&\frac{P_2(\hat \d_R)\tau_2}{\tau_1} &\dots
\\ 0&0 &\frac{\tau_3}{\tau_2}  &\frac{P_1(\hat \d_R)\tau_3}{\tau_2}&\dots
\\ 0&0 &0  &\frac{\tau_4}{\tau_3}&\dots\\ \dots &\dots &\dots &\dots&\dots
\end{array}\right),
\end{eqnarray}

  \begin{eqnarray}\label{momentR2}
\tilde \P_R^{-1}
 &=&\left(\begin{array}{ccccc}\frac{\tau_0}{\tau_1} &\frac{P_1(-\hat \d_R)\tau_1}{\tau_2} &\frac{P_2(-\hat \d_R)\tau_2}{\tau_3} &\frac{P_3(-\hat \d_R)\tau_3}{\tau_4} &\dots \\
0 &\frac{\tau_1}{\tau_2}  &\frac{P_1(-\hat \d_R)\tau_2}{\tau_3}&\frac{P_2(-\hat \d_R)\tau_3}{\tau_4} &\dots
\\ 0&0 &\frac{\tau_2}{\tau_3}  &\frac{P_1(-\hat \d_R)\tau_3}{\tau_4}&\dots
\\ 0&0 &0  &\frac{\tau_3}{\tau_4}&\dots\\ \dots &\dots &\dots &\dots&\dots
\end{array}\right).
 \end{eqnarray}
 After the following transformation
 $u_{i,j}=a_{j,j+i}$, the matrix representation of $\L$ can be expressed by $(a_{i,j})_{i,j\geq 1}$ with
\begin{equation}\label{aij}
a_{i,j}(t)=\frac{P_{i-j+N}(\hat D_L)\tau_j\circ\tau_{i-1}}{\tau_{i-1}\tau_j}
=\frac{P_{j-i+M}(\hat D_R)\tau_i\circ\tau_{j-1}}{\tau_{i-1}\tau_j}.
\end{equation}
Note here that those expressions immediately imply
\begin{align*}
a_{i,j}&=0,\qquad {\rm if}\quad j<-M+i\quad{\rm or}\quad j>N+i.
\end{align*}
That is, the Lax matrix $\L$ has the $(N,M)$ band structure.

As an example, we will  give some concrete results on $(2,2)$-BTH in the following subsection from which we can see some general patten of $(N,M)$-BTH.
\subsection{Example of the (2,2)-BTH}
Let us summarize this section by taking the (2,2)-BTH.  The Lax operator is
\begin{equation}L=\Lambda^2+u_{1}\Lambda+u_0 + u_{-1}
\Lambda^{-1}+ u_{-2}\Lambda^{-2}.
\end{equation}
Then there  will be two different fraction power of $L$, denoted as $L_N^{\frac12}$ and $L_M^{\frac12}$ respectively as following form
\begin{equation}L_N^{\frac12}=\Lambda+a_0 + a_{-1}
\Lambda^{-1}+ a_{-2}\Lambda^{-2}+\dots,
\end{equation}
\begin{equation}L_M^{\frac12}=a'_{-1}\Lambda^{-1}+a'_0 + a'_{1}
\Lambda+ a'_{2}\Lambda^{2}+\dots.
\end{equation}
We can get some relations of $\{a_i; i\leq 0\},\{ a'_j; j\geq -1\}$ with $\{u_i; -M\leq i\leq N-1\}$ as following
\begin{eqnarray}\label{a operator}
 a_0(x)&=&(1+\La)^{-1}u_1(x),\ \ \ a'_{-1}=e^{(1+\Lambda^{-1})^{-1}\log u_{-2}(x)}.
 \end{eqnarray}
Then by Lax equation, we get the $t_{2,0}$ flow of (2,2)-BTH
\begin{eqnarray}
 \partial_{2,0} L= [\Lambda +(1+\La)^{-1}u_1(x), L]
\end{eqnarray}
which correspond to
\begin{eqnarray}\label{2,0 flow}
\begin{cases}
 \partial_{2,0} u_1(x)&= u_0(x+\epsilon)-u_0(x)+u_1(x)(1-\La)(1+\La)^{-1}u_1(x)\\
 \partial_{2,0} u_0(x)&= u_{-1}(x+\epsilon)-u_{-1}(x)\\
 \partial_{2,0} u_{-1}(x)&= u_{-2}(x+\epsilon)-u_{-2}(x)+u_{-1}(x)(1-\La^{-1})(1+\La)^{-1}u_1(x)\\
  \partial_{2,0} u_{-2}(x)&=u_{-2}(x)(1-\La^{-2})(1+\La)^{-1}u_1(x).
 \end{cases}
 \end{eqnarray}
 From eqs.\eqref{2,0 flow}, we can find the equations have infinite terms because of $(1+\La)^{-1}$ which comes from the fraction power of the Lax operator. Just like the method in \cite{moleculesolutions}, for avoiding infinite sums, we use  auxiliary function $a_0(x)$ with which we can rewrite eq.\eqref{2,0 flow} as
 \begin{eqnarray}\label{2,0 flow'}
\begin{cases}
 \partial_{2,0} u_1(x)&= u_0(x+\epsilon)-u_0(x)+u_1(x)(a_0(x)-a_0(x+\ep))\\
 \partial_{2,0} u_0(x)&= u_{-1}(x+\epsilon)-u_{-1}(x)\\
 \partial_{2,0} u_{-1}(x)&= u_{-2}(x+\epsilon)-u_{-2}(x)+u_{-1}(x)(a_0(x)-a_0(x-\ep))\\
  \partial_{2,0} u_{-2}(x)&=u_{-2}(x)(a_0(x)-a_0(x-2\ep))\\
  \partial_{2,0} a_0(x+\ep)+\partial_{2,0} a_0(x)&=u_0(x+\epsilon)-u_0(x)+u_1(x)(a_0(x)-a_0(x+\ep)).
 \end{cases}
 \end{eqnarray}

 The $t_{1,0}$ flow will have finite terms as following because it does not use the fraction power of Lax operator $L$,
\begin{eqnarray}
 \partial_{1,0} L= [\Lambda^2+u_1\Lambda +u_0, L]
\end{eqnarray}
which correspond to
\begin{eqnarray}
\begin{cases}
\partial_{1,0} u_1(x)&= u_{-1}(x+2\epsilon)-u_{-1}(x)\\
 \partial_{1,0} u_0(x)&= u_{-2}(x+2\epsilon)-u_{-2}(x)+u_1(x)u_{-1}(x+\epsilon)-u_{-1}(x)u_1(x-\ep)\\
 \partial_{1,0} u_{-1}(x)&= u_1(x)u_{-2}(x+\epsilon)-u_{-2}(x)u_1(x-2\ep)+u_{-1}(x)(u_0(x)-u_0(x-\epsilon))\\
  \partial_{1,0} u_{-2}(x)&=u_{-2}(x)( u_0(x)-u_0(x-2\epsilon)).
   \end{cases}
\end{eqnarray}
For $t_{-1,0}$ flow, equations will also be complicated because of another fraction power of $L$. The equation is
\begin{eqnarray}
 \partial_{-1,0} L= -[e^{(1+\Lambda^{-1})^{-1}\log u_{-2}}\Lambda^{-1} , L]
\end{eqnarray}
which corresponds to
\begin{eqnarray}\label{-1,0 flow}
\begin{cases}
\partial_{-1,0} u_1(x)&= e^{(1+\Lambda^{-1})^{-1}\log u_{-2}(x+2\epsilon)}-e^{(1+\Lambda^{-1})^{-1}\log u_{-2}(x)}\\[1.5ex]
 \partial_{-1,0} u_0(x)&= u_1(x)e^{(1+\Lambda^{-1})^{-1}\log u_{-2}(x+\epsilon)}-e^{(1+\Lambda^{-1})^{-1}\log u_{-2}(x)}u_1(x-\ep)\\[1.5ex]
 \partial_{-1,0} u_{-1}(x)&= e^{(1+\Lambda^{-1})^{-1}\log u_{-2}(x)}(u_0(x)-u_0(x-\epsilon))\\[1.5ex]
  \partial_{-1,0} u_{-2}(x)&=u_{-1}(x)e^{(1+\Lambda^{-1})^{-1}\log u_{-2}(x-\epsilon)}-e^{(1+\Lambda^{-1})^{-1}\log u_{-2}(x)}u_{-1}(x-\epsilon).
  \end{cases}
\end{eqnarray}
From eqs.\eqref{-1,0 flow}, we can find the equations have finite terms but every term has infinite multiplication because of factor $e^{(1+\Lambda^{-1})^{-1}\log u_{-2}(x)}$ which comes from the square root  of Lax operator. Similarly for avoiding infinite multiplication, we use auxiliary function $a'_{-1}(x)$ with which we can rewrite eq.\eqref{-1,0 flow} as
\begin{eqnarray}\label{-1,0 flow'}
\begin{cases}
\partial_{-1,0} u_1(x)&= a'_{-1}(x+2\epsilon)-a'_{-1}(x)\\[1.5ex]
 \partial_{-1,0} u_0(x)&= u_1(x)a'_{-1}(x+\epsilon)-a'_{-1}(x)u_1(x-\ep)\\[1.5ex]
 \partial_{-1,0} u_{-1}(x)&= a'_{-1}(x)(u_0(x)-u_0(x-\epsilon))\\[1.5ex]
  \partial_{-1,0} u_{-2}(x)&=u_{-1}(x)a'_{-1}(x-\epsilon)-a'_{-1}(x)u_{-1}(x-\epsilon)\\
    \partial_{-1,0} a'_{-1}(x)&=u_{-1}(x).
  \end{cases}
\end{eqnarray}

Infinite sums or infinite multiplications are important properties of BTH because of nonlocal operators. For finite Lax matrix, it is not easy to construct  its fraction power. That is why we do not use fraction power of finite matrix to give Lax equations.

\sectionnew{Equivalence between $(N,M)$-BTH and $(M,N)$-BTH}\label{sec:Equivalent}
In this section, we prove that there is an equivalence between $(N,M)$-BTH and $(M,N)$-BTH in  three ways. One is to prove the equivalence in the Hirota bilinear identities basing on \cite{our}. The second one is for equivalence in specific Hirota bilinear equations. At last, we will prove the equivalence between their Lax
equations using transformation in  \cite{BlaszakMarciniakJMP}. To see the equivalence clearly, one explicit example, i.e. equivalence between $(1,2)$-BTH and $(2,1)$-BTH in Lax equations under transformation will be shown in detail.

\subsection{Equivalence in the Hirota bilinear identities}
Firstly after denoting  $\tau(x,t)$ as $\tau(x-\frac{\epsilon}{2},t)$, we  get the following
Hirota bilinear identity\cite{our}, i.e.  for
each  $m\in \Z$, $r\in \N$,
\begin{eqnarray}\notag
&& \res_{\lambda } \left\{ \lambda^{Nr+m-1}
\tau(x,t-[\lambda^{-1}]^N)\times
\tau(x-(m-1)\epsilon,t'+[\lambda^{-1}]^N) e^{\xi_L(\la, t-t')} \right\}
\\\label{HBE511x} &&=  \res_{\lambda } \left\{ \lambda^{-Mr+m-1}
\tau(x+\epsilon,t+[\lambda]^M)\times
\tau(x-m\epsilon,t'-[\lambda]^M) e^{-\xi_R(\la^{-1},t-t')}\right\} ,
\end{eqnarray}
where\begin{eqnarray*} \xi_L(\la, t)&=&\sum_{n\geq 0} \sum_
{\alpha=1}^{N}
\lambda^{N({n+1-\frac{\alpha-1}{N}})}t_{\alpha,n},\\
\xi_R(\la,t)&=&\sum_{n\geq 0} \sum_ {\beta=-M+1}^{0}
\lambda^{M({n+1+\frac{\beta}{M}})}t_{\beta,n}.
\end{eqnarray*}

The most important property that BTH has is that there are $Nr$ and $Mr$ in both sides of HBEs \eqref{HBE511x}. These two terms show the principal difference  of BTH from the two-dimensional Toda hierarchy, i.e. the constraint \eqref{two dressing}.

 Hirota bilinear identity  eq.\eqref{HBE511x}
 can lead to the following identity  under the transformation
$m\mapsto -m, x\mapsto x-m\ep, \la\mapsto \la^{-1}$
\begin{eqnarray}\notag
&& \res_{\lambda } \left\{ \la^{-1}\lambda^{-Nr+m}
\tau(x-m\ep,t-[\lambda]^N)\times
\tau(x+\epsilon,t'+[\lambda]^N) e^{\xi'_L(\la, t-t')} \right\}
\\\label{HBE511'} &&=  \res_{\lambda } \left\{ \la^{-1}\lambda^{Mr+m}
\tau(x-(m-1)\epsilon,t+[\lambda^{-1}]^M)\times
\tau(x,t'-[\lambda^{-1}]^M) e^{-\xi'_R(\la^{-1},t-t')}\right\} ,
\end{eqnarray}
where\begin{eqnarray*} \xi'_L(\la, t-t')&=&\sum_{n\geq 0} \sum_
{\alpha=1}^{N}\lambda^{-N({n+1-\frac{\alpha-1}{N}})}(t_{\alpha,n}-t'_{\alpha,n}),\\
\xi'_R(\la^{-1},t-t')&=&\sum_{n\geq 0} \sum_ {\beta=-M+1}^{0}\lambda^{M({n+1+\frac{\beta}{M}})}(t_{\beta,n}-t'_{\beta,n}).
\end{eqnarray*}
The eq.\eqref{HBE511'} can be rewritten as following identity after the interchanging of $t$ and $t'$
\begin{eqnarray}\notag
&& \res_{\lambda } \left\{ \la^{-1}\lambda^{-Nr+m}
\tau(x+\epsilon,t+[\lambda]^N)\times\tau(x-m\ep,t'-[\lambda]^N)
 e^{\xi'_L(\la,t'-t)} \right\}
\\\label{HBE511111} &&=  \res_{\lambda } \left\{ \la^{-1}\lambda^{Mr+m}
\tau(x,t-[\lambda^{-1}]^M) \times \tau(x-(m-1)\epsilon,t'+[\lambda^{-1}]^M)
e^{-\xi'_R(\la^{-1},t'-t)}\right\}
\end{eqnarray}
which can be further  rewritten as following
\begin{eqnarray}\notag
&&  \res_{\lambda } \left\{ \la^{-1}\lambda^{Mr+m}
\tau(x,t-[\lambda^{-1}]^M) \times \tau(x-(m-1)\epsilon,t'+[\lambda^{-1}]^M)
e^{\xi_R(\la, t-t')}\right\}
\\\label{HBE5111'} &&= \res_{\lambda } \left\{ \la^{-1}\lambda^{-Nr+m}
\tau(x+\epsilon,t+[\lambda]^N)\times\tau(x-m\ep,t'-[\lambda]^N)
 e^{-\xi_L(\la^{-1},t-t')} \right\}.
\end{eqnarray}
Eq.\eqref{HBE5111'} is obviously $(M,N)$-BTH comparing to eq.\eqref{HBE511x} if we change time variable $t_{\gamma,n}$ to $t_{1-\gamma,n}$, i.e.
subscript $L\leftrightarrow R$. Therefore for $(M,N)$-BTH, Eq.\eqref{HBE5111'} is in fact changed into  the following equation under
transformation  $t_{\gamma,n}\rightarrow t_{1-\gamma,n}$
\begin{eqnarray}\notag
&&  \res_{\lambda } \left\{ \la^{-1}\lambda^{Mr+m}
\tau(x,t-[\lambda^{-1}]^M) \times \tau(x-(m-1)\epsilon,t'+[\lambda^{-1}]^M)
e^{\xi_L(\la, t-t')}\right\}
\\\label{HBE5111'1} &&= \res_{\lambda } \left\{ \la^{-1}\lambda^{-Nr+m}
\tau(x+\epsilon,t+[\lambda]^N)\times\tau(x-m\ep,t'-[\lambda]^N)
 e^{-\xi_R(\la^{-1},t-t')} \right\}.
\end{eqnarray}
Because transforms $m\mapsto -m, x\mapsto x-m\ep, \la\mapsto \la^{-1}$ do not change the equation itself. Therefore
we can say that the $(N,M)$-BTH is equivalent to the $(M,N)$-BTH under the transformation
$t_{\gamma,n}$ $\rightarrow$ $t_{1-\gamma,n}$.

\subsection{Equivalence in the Hirota bilinear equations}
The bilinear identity eq.\eqref{HBE511x} of the BTH  can be equivalently expressed as \cite{our}
\begin{align}\nonumber
 &\res_{\lambda } \left\{ \lambda^{Nr+m-1}
\tau_{j-(m-1)}(t+y+[\lambda^{-1}]^N)\, \tau_j(t-y-[\lambda^{-1}]^N)\,
e^{\xi_L(\la,-2y)} \right\}
\\\label{HBE5111}
&=  \res_{\lambda } \left\{ \lambda^{-Mr+m-1}
\tau_{j+1}(t-y+[\lambda]^M)\,
\tau_{j-m}(t+y-[\lambda]^M) \,e^{\xi_R(\la^{-1},2y)}\right\}.
\end{align}

To be specific,  we will give  the equivalence from the concrete Hirota equations between $(N,M)$-BTH and $(M,N)$-BTH as following which will
also be used to derive specific primary Hirota equations of BTH in the next section. \\
For $(N,M)$-BTH,
in term with $\prod y^{k_1}_{\alpha_1,l_1}y^{k_2}_{\alpha_2,l_2}\dots y^{k_s}_{\alpha_s,l_s}\prod y^{k'_1}_{\beta_1,l'_1}y^{k'_2}_{\beta_2,l'_2}\dots y^{k'_p}_{\beta_p,l'_p}, -M+1\leq\beta_i\leq 0, 1\leq\alpha_i\leq N$, the
Hirota equation is as

\begin{eqnarray}\notag
&&\prod_{v=1}^s \frac{(-D_{\alpha_v,l_v})^{k_v}}{k_v!}(\sum_{i'_{1}=0}^{k'_1}\dots\sum_{i'_{p}=0}^{k'_p}\prod_{q=1}^p\frac{(-D_{\beta_{q},l'_{q}})^{i'_{q}}}{i'_{q}!}
\frac{2^{k'_q-i'_q}}{(k'_q-i'_q)!}
P_{\sum_{q=1}^p M(l'_q+1+\frac{\beta_q}{M})(k'_q-i'_q)+Mr-m}(\hat D_R))\tau_{n+1}\tau_{n-m}\\ \notag
&&=\prod_{v=1}^p \frac{D^{k'_v}_{\beta_v,l'_v}}{k'_v!}
(\sum_{i_{1}=0}^{k_1}\dots\sum_{i_{s}=0}^{k_s}\prod_{q=1}^s\frac{(D_{\alpha_{q},l_{q}})^{i_{q}}}{i_{q}!}
\frac{(-2)^{k_q-i_q}}{(k_q-i_q)!}
P_{\sum_{q=1}^s N(l_q+1-\frac{\alpha_q-1}{N})(k_q-i_q)+Nr+m}(\hat D_L))
\tau_{n-m+1}\tau_{n},\\\label{NMmixed}
\end{eqnarray}
where $P_{k}$ are Schur polynomial as defined in \eqref{shurfunction}.
After the transformation
$m\mapsto -m, n\mapsto n-m,$ the identity eq.\eqref{NMmixed} becomes

\begin{eqnarray}\notag
&&\prod_{v=1}^p \frac{D^{k'_v}_{\beta_v,l'_v}}{k'_v!}
(\sum_{i_{1}=0}^{k_1}\dots\sum_{i_{s}=0}^{k_s}\prod_{q=1}^s\frac{(D_{\alpha_{q},l_{q}})^{i_{q}}}{i_{q}!}
\frac{(-2)^{k_q-i_q}}{(k_q-i_q)!}
P_{\sum_{q=1}^s N(l_q+1-\frac{\alpha_q-1}{N})(k_q-i_q)+Nr-m}(\hat D_L))\tau_{n+1}\tau_{n-m}\\ \notag
&&=\prod_{v=1}^s \frac{(-D_{\alpha_v,l_v})^{k_v}}{k_v!}(\sum_{i'_{1}=0}^{k'_1}\dots\sum_{i'_{p}=0}^{k'_p}\prod_{q=1}^p\frac{(-D_{\beta_{q},l'_{q}})^{i'_{q}}}{i'_{q}!}
\frac{2^{k'_q-i'_q}}{(k'_q-i'_q)!}
P_{\sum_{q=1}^p M(l'_q+1+\frac{\beta_q}{M})(k'_q-i'_q)+Mr+m}(\hat D_R))\\ \label{NMmixed'1}
&&\tau_{n-m+1}\tau_{n}.
\end{eqnarray}
After  doing the transformation $D_{\gamma, l}=D'_{1-\gamma, l}, -M+1\leq\gamma\leq N$, eq.\eqref{NMmixed'1} becomes

\begin{eqnarray}\notag
&&\prod_{v=1}^p \frac{(D'_{1-\beta_v,l'_v})^{k'_v}}{k'_v!}(\sum_{i_{1}=0}^{k_1}\dots\sum_{i_{s}=0}^{k_s}\prod_{q=1}^s\frac{(D'_{1-\alpha_{q},l_{q}})^{i_{q}}}
{i_{q}!}
\frac{(-2)^{k_q-i_q}}{(k_q-i_q)!}
P_{\sum_{q=1}^s N(l_q+1+\frac{1-\alpha_q}{N})(k_q-i_q)+Nr-m}(\hat D'_R))\\
\notag &&\tau_{n+1}\tau_{n-m}=\prod_{v=1}^s \frac{(-D'_{1-\alpha_v,l_v})^{k_v}}{k_v!}
\\
&&(\sum_{i_{1}=0}^{k'_1}\dots\sum_{i_{p}=0}^{k'_p}\prod_{q=1}^p\frac{(-D'_{1-\beta_{q},l'_{q}})^{i'_{q}}}{i'_{q}!}
\frac{2^{k'_q-i'_q}}{(k'_q-i'_q)!}
P_{\sum_{q=1}^p M(l'_q+1+\frac{\beta}{M})(k'_q-i'_q)+Mr+m}(\hat D'_L))\tau_{n-m+1}\tau_{n}.
\end{eqnarray}
which is the  term with $\prod y^{k'_1}_{1-\beta_1,l'_1}y^{k'_2}_{1-\beta_2,l'_2}\dots y^{k'_p}_{1-\beta_p,l'_p}\prod y^{k_1}_{1-\alpha_1,l_1}y^{k_2}_{1-\alpha_2,l_2}\dots y^{k_s}_{1-\alpha_s,l_s}, 1\leq1-\beta_i\leq M, -N+1\leq1-\alpha_i\leq 0$
for $(M,N)$-BTH.
So there is a correspondence of $(N,M)$-BTH and $(M,N)$-BTH in Hirota equations under the meaning of following  derivatives' correspondence
$D'_{\gamma, l}\leftrightarrow D_{1-\gamma, l}, -M+1\leq\gamma\leq N$, i.e. $D'_L\leftrightarrow D_R, D'_R\leftrightarrow D_L$.

Because $(N,M)$-BTH and $(M,N)$-BTH are equivalent, we can only consider BTH in  the case of $N\leq M$ later.

\subsection{Equivalence in the Lax equations}
Using the gauge transformation and linear transformation mentioned in \cite{BlaszakMarciniakJMP}, we can prove the equivalence between
 $(N,M)$-BTH and $(M,N)$-BTH under the meaning of Lax equations.

Firstly we need to introduce the following proposition
basing on Theorem 6 in \cite{BlaszakMarciniakJMP}.
\begin{proposition}\label{gauge}
If $\L$ satisfies Lax equations \eqref{BTHLax} and
let  $\Phi=\Phi(u,t)$ satisfies condition
\begin{eqnarray}\label{phi}
\d_{t_{\gamma,n}}\Phi= (B_{\gamma,n})_0\Phi,  \ \  -M+1\leq\gamma\leq N, n \geq 0,
\end{eqnarray}
(where subscript $0$ denotes the projection to term of $\La^0$), then $\tilde \L=\Phi^{-1}\L \Phi$ satisfies the hierarchy
\begin{eqnarray}
\d_{t_{\gamma,n}}\tilde\L= [(\tilde B_{\gamma,n})_{\geq 1}, \tilde\L], \ \  -M+1\leq\gamma\leq N, n \geq 0,
\end{eqnarray}
where
\begin{eqnarray}
\tilde B_{\gamma,n}=\Phi^{-1}B_{\gamma,n} \Phi.
\end{eqnarray}
\end{proposition}
\begin{proof}
Because of \eqref{BTHLax} and \eqref{phi}, the following calculation holds
\begin{eqnarray*}
\d_{t_{\gamma,n}}\tilde\L- [(\tilde B_{\gamma,n})_{\geq 1}, \tilde\L]&=&\d_{t_{\gamma,n}}(\Phi^{-1}\L \Phi)-\Phi^{-1}[(B_{\gamma,n})_{\geq 1},\L]\Phi\\
&=&\Phi^{-1}(\d_{t_{\gamma,n}}\L- [(B_{\gamma,n})_+, \L])\Phi-[\Phi^{-1}(\d_{t_{\gamma,n}}\Phi-(B_{\gamma,n})_0\Phi), \tilde\L]\\
&=&0.
\end{eqnarray*}
Then we finished the proof of this proposition.
\end{proof}
Proposition \ref{gauge} tells us that the gauge
transformation from Theorem 6 in \cite{BlaszakMarciniakJMP} can be extended on the fractional powers
of Lax operators.
 Now let us introduce the following notation for anti-involution map $``\dag"$:
 If $\L$ is as form \eqref{LBTH}, then
  \begin{eqnarray*}\label{LBTH1}
\L^\dag:&=&\Lambda^{M}u_{-M}(x)
+\Lambda^{M-1}u_{-M+1}(x)+\cdots+\Lambda^{-N+1}u_{N-1}(x)+\Lambda^{-N}\\
&=&u_{-M}(x+M\ep)
\Lambda^{M}+u_{-M+1}(x+(M+1)\ep)\Lambda^{M-1}+\cdots+u_{N-1}(x-(N-1)\ep)\Lambda^{-N+1}+\Lambda^{-N}.
\end{eqnarray*}
 By calculation, one can prove the following two lemmas similar as \cite{BlaszakMarciniakJMP}.

 \begin{lemma} For $-M+1\leq\gamma\leq N, n \geq 0$, following identity holds for any integer $k$
 \begin{eqnarray}
(( B_{\gamma,n})_{\geq k})^{\dag}=((B_{\gamma,n})^{\dag})_{\leq -k}.
\end{eqnarray}
 \end{lemma}
 Using above lemma, we can prove the following lemma directly.
  \begin{lemma} \label{01switch}
$\d_{t_{\gamma,n}}\L= [( B_{\gamma,n})_{\geq 0}, \L]$ can lead to $\d_{t_{\gamma,n}}\L^{\dag}= [((B_{\gamma,n})^{\dag})_{\geq 1}, \L^{\dag}]$
and
$\d_{t_{\gamma,n}}\L= [( B_{\gamma,n})_{\geq 1}, \L]$ can lead to $\d_{t_{\gamma,n}}\L^{\dag}= [((B_{\gamma,n})^{\dag})_{\geq 0}, \L^{\dag}].$
 \end{lemma}
 By above two lemmas and Proposition \ref{gauge}, we can prove the following important theorem.
 \begin{theorem}
 Lax equation \eqref{BTHLax} of $(N,M)$-BTH with Lax operator
 \begin{eqnarray}\L_{N,M}=\Lambda^{N}+u_{N-1}\Lambda^{N-1}+\cdots+u_0+\cdots + u_{-M}\La^{-M}\end{eqnarray}
 is equivalent to  $(M,N)$-BTH with  Lax operator
 \begin{eqnarray}\L_{M,N}=\Lambda^{M}+\tilde u_{M-1}\Lambda^{M-1}+\cdots+\tilde u_0+\cdots + \tilde u_{-N}\La^{-N}\end{eqnarray}
 under  Miura map: $\tilde u_j(x,t)= u_{-j}(x+j\ep)e^{\frac{1-\La^j}{1-\La^{-M}}u_{-M}(x,t)}.$
 \end{theorem}
 \begin{proof}
 For Lax operator $\L_{N,M}=\Lambda^{N}+u_{N-1}\Lambda^{N-1}+\cdots+u_0+\cdots + u_{-M}\La^{-M}$  of $(N,M)$-BTH which satisfies Lax equations \eqref{BTHLax},
one can choose  $\Psi=\Psi(u,t)=e^{(1-\La^{-M})^{-1}u_{-M}(x,t)}$ which satisfies condition
\begin{eqnarray}\label{Psi}
\d_{t_{\gamma,n}}\Psi= (B_{\gamma,n})_0\Psi,  \ \  -M+1\leq\gamma\leq N, n \geq 0.
\end{eqnarray}
Then $\tilde \L_{N,M}=\Psi^{-1}\L_{N,M} \Psi=\bar u_N\Lambda^{N}+\bar u_{N-1}\Lambda^{N-1}+\cdots+\bar u_0+\cdots + \La^{-M}$ satisfies the hierarchy
\begin{eqnarray}
\d_{t_{\gamma,n}}\tilde\L_{N,M}= [(\tilde B_{\gamma,n})_{\geq 1}, \tilde\L_{N,M}], \ \  -M+1\leq\gamma\leq N, n \geq 0,
\end{eqnarray}
where
\begin{eqnarray}
\tilde B_{\gamma,n}=\Psi^{-1}B_{\gamma,n} \Psi, \ \ \bar u_i(x)=\Psi^{-1}(x)u_i(x) \Psi(x+i\ep).
\end{eqnarray}
Using Lemma \ref{01switch}, one can derive
\begin{eqnarray}
\d_{t_{\gamma,n}}\tilde\L_{N,M}^{\dag}= [((\tilde B_{\gamma,n})^{\dag})_{\geq 0}, \tilde\L_{N,M}^{\dag}], \ \  -M+1\leq\gamma\leq N, n \geq 0,
\end{eqnarray}
One can choose Lax operator $\L_{M,N}=\tilde\L_{N,M}^{\dag}=\Lambda^{M}+\tilde u_{M-1}\Lambda^{M-1}+\cdots+\tilde u_0+\cdots + \tilde u_{-N}\La^{-N}$  of $(M,N)$-BTH
 with $$\tilde u_j(x)=\bar u_{-j}(x+j\ep)=\Psi^{-1}(x+j\ep)u_{-j}(x+j\ep) \Psi(x)=
 u_{-j}(x+j\ep)e^{\frac{1-\La^j}{1-\La^{-M}}u_{-M}(x,t)}.$$
 The $(M,N)$-BTH with this Lax matrix will be equivalent to the original $(N,M)$-BTH.
  \end{proof}
To be illustrative, we will give some simplest concrete examples of
equivalence in the next subsection which includes nonlocal flows of $(2,1)$-BTH and
$(1,2)$-BTH. Here we will also consider the interpolated form of BTH. At this time, we will use a powerful tool called gauge transformation to prove that equivalence.

\subsection{Equivalence between $(1,2)$-BTH and $(2,1)$-BTH}

Using Proposition \ref{gauge}, we will see the equivalence between $(1,2)$-BTH and $(2,1)$-BTH in detail.
Here we only give the primary flows of them to see the equivalence.

{\bf $(1,2)$-BTH:} The Lax operator of  $(1,2)$-BTH is as following
\begin{eqnarray}
\L_{1,2}= \La+u_0+u_{-1}\La^{-1}+u_{-2}\La^{-2}.
\end{eqnarray}
 $(1,2)$-BTH  has the following primary equations
\begin{eqnarray}
\partial_{1,0} \L_{1,2}= [\Lambda +u_0, \L_{1,2}],
\end{eqnarray}
and
\begin{eqnarray}
 \partial_{-1,0} \L_{1,2}= -[e^{(1+\Lambda^{-1})^{-1}\log u_{-2}}\Lambda^{-1} , \L_{1,2}],
\end{eqnarray}
which further lead to
\begin{eqnarray}\label{-1,0flow}
\begin{cases}
\partial_{-1,0} u_0(x)= e^{(1+\Lambda^{-1})^{-1}\log u_{-2}(x+\epsilon)}-e^{(1+\Lambda^{-1})^{-1}\log u_{-2}(x)},\\
\partial_{-1,0} u_{-1}(x)= e^{(1+\Lambda^{-1})^{-1}\log u_{-2}(x)}(u_0(x)-u_0(x-\epsilon)),\\
  \partial_{-1,0} u_{-2}(x)=u_{-1}(x)e^{(1+\Lambda^{-1})^{-1}\log u_{-2}(x-\epsilon)}-e^{(1+\Lambda^{-1})^{-1}\log u_{-2}(x)} u_{-1}(x-\epsilon).
  \end{cases}
\end{eqnarray}
and
\begin{eqnarray}\label{1,0flow}
\begin{cases}
\partial_{1,0} u_0(x)= u_{-1}(x+\epsilon)-u_{-1}(x),\\
\partial_{1,0} u_{-1}(x)= u_{-2}(x+\epsilon)-u_{-2}(x)+u_{-1}(x)(u_0(x)-u_0(x-\epsilon)),\\
 \partial_{1,0} u_{-2}(x)=u_{-2}(x)( u_0(x)-u_0(x-2\epsilon)).
  \end{cases}
\end{eqnarray}

{\bf $(2,1)$-BTH:} The Lax operator of $(2,1)$-BTH is as following
\begin{eqnarray}
\L_{2,1}= \La^{2}+\bar u_1\La+\bar u_0+\bar u_{-1}\La^{-1}.
\end{eqnarray}
The equations \eqref{BTHLax}in this case are as follows
\begin{eqnarray}
&& \partial_{2,0} \L_{2,1}= [\Lambda +(1+\La)^{-1}\bar u_1(x), \L_{2,1}],
\\
&&\partial_{1,0} \L_{2,1}= [\Lambda^2+\bar u_1\Lambda +\bar u_0, \L_{2,1}],
\end{eqnarray}
which further lead to the following concrete equations
\begin{eqnarray}\label{N=2,M=12,0flow}
\begin{cases}
 \partial_{2,0} \bar u_1(x)= \bar u_1(x+\epsilon)-\bar u_1(x)+\bar u_1(x)(1-\La)(1+\La)^{-1}\bar u_1(x),\\
\partial_{2,0} \bar u_0(x)= \bar u_{-1}(x+\epsilon)-\bar u_{-1}(x),\\
\partial_{2,0} \bar u_{-1}(x)= \bar u_{-1}(x)(1-\La^{-1})(1+\La)^{-1}\bar u_1(x),
 \end{cases}
\end{eqnarray}

\begin{eqnarray}\label{N=2,M=11,0flow}
\begin{cases}
\partial_{1,0} \bar u_1(x)= \bar u_{-1}(x+2\epsilon)-\bar u_{-1}(x),\\
\partial_{1,0} \bar u_0(x)= \bar u_{-2}(x+2\epsilon)-\bar u_{-2}(x)+\bar u_1(x)\bar u_{-1}(x+\epsilon)-\bar u_{-1}(x)\bar u_1(x-\ep),\\
\partial_{1,0} \bar u_{-1}(x)= \bar u_{-1}(x)(\bar u_0(x)-\bar u_0(x-\epsilon)).
 \end{cases}
\end{eqnarray}
It seems that eq.\eqref{N=2,M=12,0flow} and eq.\eqref{N=2,M=11,0flow} are quite different from
eq.\eqref{-1,0flow} and eq.\eqref{1,0flow} respectively. In fact after doing gauge transformation on $(2,1)$-BTH as following, we
can find the equivalent relation between these equations.

Now we consider function $\phi$ has form
\begin{eqnarray}
 \phi=e^{(1-\Lambda)^{-1}\log \bar u_{-1}(x)},
\end{eqnarray}
then $\hat\L_{2,1}:= \phi^{-1}\L_{2,1}\phi$ will have form
\begin{eqnarray}
 \hat\L_{2,1}= v_2\La^2+v_1\La+v_0+\La^{-1}.
\end{eqnarray}
The relation of $v_i (0\leq i \leq 2)$ and $\bar u_i (-1\leq i \leq 1)$ are like
\begin{eqnarray}
 v_2&=& \phi^{-1}\phi(x+2\ep),\ \  v_1= \phi^{-1}u_1\phi(x+\ep),\ \
  v_0= u_0.
\end{eqnarray}
Therefore we get following new flows on new Lax operator $\hat\L_{2,1}$ using Proposition \ref{gauge}
\begin{eqnarray}
&& \partial_{2,0} \hat\L_{2,1}= [e^{(1+\Lambda)^{-1}\log v_2(x)}\Lambda,\hat \L_{2,1}],
\\
&&\partial_{1,0} \hat\L_{2,1}= [v_2\La^2+v_1\La^1, \hat\L_{2,1}],
\end{eqnarray}
which further leads to
\begin{eqnarray}\label{2,0flow1}
\begin{cases}
\partial_{2,0} v_0(x)= e^{(1+\Lambda)^{-1}\log v_{2}(x)}-e^{(1+\Lambda)^{-1}\log v_{2}(x-\ep)},\\
\partial_{2,0} v_{1}(x)= e^{(1+\Lambda)^{-1}\log v_{2}(x)}(v_0(x+\ep)-v_0(x)),\\
  \partial_{2,0} v_{2}(x)=v_{1}(x+\ep)e^{(1+\Lambda)^{-1}\log v_{2}(x)}-e^{(1+\Lambda)^{-1}\log v_{2}(x+\ep)} v_{1}(x).
  \end{cases}
\end{eqnarray}
and
\begin{eqnarray}\label{1,0flow1}
\begin{cases}
\partial_{1,0} v_0(x)= v_{1}(x)-v_{1}(x-\ep),\\
\partial_{1,0} v_{1}(x)= v_{2}(x)-v_{2}(x-\ep)+v_{1}(x)(v_0(x+\ep)-v_0(x)),\\
 \partial_{1,0} v_{2}(x)=v_{2}(x)( v_0(x+2\ep)-v_0(x)).
  \end{cases}
\end{eqnarray}
Comparing  eq.\eqref{2,0flow1}, eq.\eqref{1,0flow1} with eq.\eqref{-1,0flow} and eq.\eqref{1,0flow}, we can find these two pairs of flows are equivalent under Miura map $ u_j=v_{-j}(x+j\ep)(0\leq j\leq 2)$.
This proves the equivalence between $(1,2)$-BTH and $(2,1)$-BTH.

Using bilinear identities got by comparing every term of Hirota bilinear identities of BTH in this section, in the next section we will derive all primary Hirota equations of BTH to see its inner structure.

\sectionnew{Hirota equations and solutions of the BTH}\label{sec:Hirota}

Hirota bilinear equations(HBEs) are central object in Sato theory. From HBEs, we can derive the structure of solution.
This is a great motivation for us to consider HBEs of the BTH.
The Hirota bilinear equations of the BTH can be derived from \eqref{HBE5111} which comes from HBEs in \cite{our}. In particular, the following proposition will list all  the Hirota equations
for the primary variables, i.e. $t_{\gamma,n}$ with $n=0$.
\begin{proposition}\label{primary equaitons}
For $(N,M)$-BTH, we have the following identities for primary derivatives which are equivalent to all the
primary Hirota equations.
\begin{align}\label{derivative hirotaM1}
&\left(D_{\beta,0}-P_{M+\beta}(\hat D_R)\right)\tau_{n+1}\circ\tau_{n}=0,\\\label{KP hirotaM}
&\left(D_{\beta,0}D_{-M+1,0}-2P_{M+\beta+1}(\hat D_R)\right)\tau_{n}\circ\tau_{n}=0,\\\label{2D Toda hirotaN}
&D_{\beta,0}D_{N,0}\tau_{n}\circ\tau_{n}=2P_{M+\beta-1}(\hat D_R)\tau_{n+1}\circ\tau_{n-1},
\\[1.0ex]
\label{2D Toda hirotaM}
&D_{\alpha,0}D_{-M+1,0}\tau_{n}\circ\tau_{n}=2P_{N-\alpha}(\hat D_L)\tau_{n+1}\circ\tau_{n-1}\\
\label{KP hirotaN}
&\left(D_{\alpha,0}D_{N,0}-2P_{N-\alpha+2}(\hat D_L)\right)\tau_{n}\circ\tau_{n}=0,\\
\label{derivative hirotaN}
&\left(D_{\alpha,0}-P_{N-\alpha+1}(\hat D_L)\right)\tau_{n+1}\circ\tau_{n}=0.
\end{align}
where $P_{k}$ are Schur polynomial as defined in \eqref{shurfunction}.
\end{proposition}
\begin{proof}
In eq.\eqref{HBE5111}, for term  $y^k_{\beta,l}, -M+1\leq\beta\leq 0$ in $(N,M)$-BTH, following Hirota equation holds

\begin{eqnarray}\notag
(\sum_{i=0}^k\frac{(-D_{\beta,l})^i}{i!}\frac{2^{k-i}}{(k-i)!}
P_{M(l+1+\frac{\beta}{M})(k-i)+Mr-m}(\hat D_R))\tau_{n+1}\tau_{n-m}=\frac{(D_{\beta,l})^k}{k!}P_{Nr+m}(\hat D_L)\tau_{n-m+1}\tau_{n},\\
\label{hirota for NM alpha}
\end{eqnarray}
which can also be got from eq.\eqref{NMmixed}.\\
For $y^k_{\alpha,l}, 1\leq\alpha\leq N$ in $(N,M)$-BTH, following Hirota equation holds

\begin{eqnarray}\notag
(\sum_{i=0}^k\frac{(-D_{\alpha,l})^i}{i!}\frac{2^{k-i}}{(k-i)!}
P_{N(l+1-\frac{\alpha-1}{N})(k-i)+Nr+m}(\hat D_L))\tau_{n-m+1}\tau_{n}=\frac{(D_{\alpha,l})^k}{k!}P_{Mr-m}(\hat D_R)\tau_{n+1}\tau_{n-m}.\\
\label{hirota for NM beta}
\end{eqnarray}

For $y_{\beta,0}, -M+1\leq\beta\leq 0$ in $(N,M)$-BTH, the following Hirota equation holds

\begin{eqnarray}\notag
(\sum_{i=0}^1\frac{(-D_{\beta,l})^i}{i!}\frac{2^{1-i}}{(1-i)!}
P_{M(l+1+\frac{\beta}{M})(1-i)+Mr-m}(\hat D_R))\tau_{n+1}\tau_{n-m}=D_{\beta,0}P_{Nr+m}(\hat D_L)\tau_{n-m+1}\tau_{n},
\end{eqnarray}
which is
\begin{eqnarray}\notag
(2
P_{M(1+\frac{\beta}{M})+Mr-m}(\hat D_R)-D_{\beta,0}P_{Mr-m}(\hat D_R))\tau_{n+1}\tau_{n-m}=D_{\beta,0}P_{Nr+m}(\hat D_L)\tau_{n-m+1}\tau_{n}.
\end{eqnarray}
Set $r=0, m=-1,$
for term with $y_{\beta,0}, -M+1\leq\beta\leq 0$,
we get
\begin{eqnarray}\notag
(2
P_{M+\beta+1}(\hat D_R)-D_{\beta,0}D_{-M+1,0})\tau_{n+1}\tau_{n+1}=0.
\end{eqnarray}
When $r=0, m=0$,\\
for term with $y_{\beta,0}, -M+1\leq\beta\leq 0$ in $(N,M)$-BTH,
we get
\begin{eqnarray}\label{relations M}
(
P_{M+\beta}(\hat D_R)-D_{\beta,0})\tau_{n+1}\tau_{n}=0.
\end{eqnarray}

Set $r=0, m=1$,
for term $y_{\beta,0}, -M+1\leq\beta\leq 0$ in $(N,M)$-BTH, we get

\begin{eqnarray}\notag
2
P_{M+\beta-1}(\hat D_R)\tau_{n+1}\tau_{n-1}=D_{\beta,0}D_{N,0}\tau_{n}\tau_{n}.
\end{eqnarray}

For term $y_{\alpha,0}, 1\leq\alpha\leq N$ in $(N,M)$-BTH, following Hirota equation holds

\begin{eqnarray}\notag
(\sum_{i=0}^1\frac{(-D_{\alpha,l})^i}{i!}\frac{2^{1-i}}{(1-i)!}
P_{N(1-\frac{\alpha-1}{N})(1-i)+Nr+m}(\hat D_L))\tau_{n-m+1}\tau_{n}=D_{\alpha,0}P_{Mr-m}(\hat D_R)\tau_{n+1}\tau_{n-m},\\
\label{hirota for NM beta'}
\end{eqnarray}

which is

\begin{eqnarray}\notag
(2
P_{N(1-\frac{\alpha-1}{N})+Nr+m}(\hat D_L)-D_{\alpha,0}P_{Nr+m}(\hat D_L))\tau_{n-m+1}\tau_{n}=D_{\alpha,0}P_{Mr-m}(\hat D_R)\tau_{n+1}\tau_{n-m}.
\end{eqnarray}
For $r=0, m=-1$, it further leads to
\begin{eqnarray}\notag
2
P_{N-\alpha}(\hat D_L)\tau_{n+2}\tau_{n}=D_{\alpha,0}D_{-M+1,0}\tau_{n+1}\tau_{n+1}.
\end{eqnarray}

Set  $r=0, m=0$,  for term with $y_{\alpha,0}, 1\leq\alpha\leq N$ in $(N,M)$-BTH,
following equation succeeds
\begin{eqnarray}\notag
(
P_{N-(\alpha-1)}(\hat D_L)-D_{\alpha,0})\tau_{n+1}\tau_{n}=0
\end{eqnarray}
Set $r=0, m=1$, we get equation
\begin{eqnarray}\notag
(
\frac12D_{\alpha,0}D_{N,0}-P_{N-\alpha+2}(\hat D_L))\tau_{n}\tau_{n}=0.
\end{eqnarray}

You can get the primary Hirota equations from the value of $k=0,1$. The other values of $k$ will give higher order derivatives of the hierarchy.

When $k=0$, the equations eq.\eqref{hirota for NM alpha} and eq.\eqref{hirota for NM beta}will give

\begin{eqnarray}\label{hirota for NM beta0}
P_{Nr+m}(\hat D_L)\tau_{n-m+1}\tau_{n}=P_{Mr-m}(\hat D_R)\tau_{n+1}\tau_{n-m}.
\end{eqnarray}
when $r=1, m=0$, eq.\eqref{hirota for NM beta0} becomes

\begin{eqnarray}\label{hirota for NM beta01}
P_{N}(\hat D_L)\tau_{n+1}\tau_{n}=P_{M}(\hat D_R)\tau_{n+1}\tau_{n}.
\end{eqnarray}

when $r=1, m=-1$, eq.\eqref{hirota for NM beta0} becomes

\begin{eqnarray}\label{hirota for NM beta02}
P_{N-1}(\hat D_L)\tau_{n+2}\tau_{n}=P_{M+1}(\hat D_R)\tau_{n+1}\tau_{n+1}.
\end{eqnarray}

when $r=1, m=1$, we get

\begin{eqnarray}\label{hirota for NM beta03}
P_{N+1}(\hat D_L)\tau_{n}\tau_{n}=P_{M-1}(\hat D_R)\tau_{n+1}\tau_{n-1}.
\end{eqnarray}
The three equations eq.\eqref{hirota for NM beta01}, eq.\eqref{hirota for NM beta02}, eq.\eqref{hirota for NM beta03}
can  be derived from eq.\eqref{derivative hirotaM1}-eq.\eqref{2D Toda hirotaM}.
These equations discussed above are all the primary Hirota equations. The other values of $r$ and $m$ will include higher derivatives of the hierarchy.
\end{proof}
From these primary Hirota equations, we can see BTH has ample structure information. We can say BTH contains discrete KP equation(from eq.\eqref{KP hirotaM} and eq.\eqref{KP hirotaN}), NLS equations(from eq.\eqref{derivative hirotaM1} and eq.\eqref{derivative hirotaN} ) and 2-dimensional Toda lattice equation(from eq.\eqref{2D Toda hirotaN} and eq.\eqref{2D Toda hirotaM}).  Comparing with Hirota equations of  the two-dimensional Toda hierarchy we find BTH have more constraints on equations which comes from the equivalence of $\d_{t_{1,n}}$ and $\d_{t_{0,n}}$. These information help us to get the solution of BTH which will be given in the next subsection. From all the primary Hirota equations mentioned above, we can get that the solution of BTH should have double-wronskian structure\cite{molecule solutions}. In paper \cite{molecule solutions}, if we impose the vanishing of $\d_y$ derivatives on tau functions the molecule equation in fact becomes our  $(2,1)$-BTH.  Using the same method in \cite{molecule solutions}, we  proved the  double-wronskian solution structure satisfy all the primary Hirota equations of $(N,M)$-BTH. But later we find there is another much simpler way to get the structure naturally which will be mentioned in the next subsection.  So we prefer this simpler way  to the way in \cite{molecule solutions}.

\subsection{Solutions of the BTH in the semi-infinite matrix}
Now we construct tau function for the BTH in the semi-infinite matrix representation, that is,
$(a_{i,j})_{ i,j\ge 1}$.
In order to do this, we first introduce the wave operators $\W_L$ and $\W_R$ associated with
the dressing operators $\P_L$ and $\P_R$,
\begin{align}\label{W_L}
 \W_L(x,t,\Lambda) &= \P_L(x,t,\Lambda)\circ
 \exp\left({
\sum_{n\geq 0} \sum_ {\alpha=1}^{N}
\Lambda^{N({n+1-\frac{\alpha-1}{N}})}t_{\alpha, n}  }\right),\\ \label{W_R}
\W_R(x,t,\Lambda)& =\P_R(x,t,\Lambda)\circ \exp\left(- \sum_{n\geq 0} \sum_
{\beta=-M+1}^{0}
\Lambda^{-M({n+1+\frac{\beta}{M}})}t_{\beta,
n}\right).
 \end{align}
 By sato equations \eqref{bn1}, we can have following identities \cite{our}
 \begin{equation} \notag
 \partial_{\alpha ,
n}\W_{L} :=
\begin{cases}
(B_{\alpha,n})_{+}\W_{L},&\alpha=N\dots 1,\\
 -(B_{
\alpha,n})_{-}\W_{L}, &\alpha = 0\dots -M+1,
\end{cases}
\end{equation}
\begin{equation} \notag
\partial_{\alpha,n}\W_{R}:=
\begin{cases}
(B_{\alpha,n})_{+}\W_{R}, &\alpha=N\dots 1,\\
 -(B_{\alpha,n})_{-}\W_{R}, &\alpha = 0\dots -M+1.
  \end{cases}
\end{equation}
 One can then prove  that the product $\W_L^{-1}\W_R$ is invariant under all the flows, i.e.
\begin{equation} \notag
\partial_{\gamma,n}(\W^{-1}_{L} \W_{R})=0.
\end{equation}
Therefore
\begin{equation} \notag
\W^{-1}_{L} \W_{R}(x,t,\Lambda)=\W^{-1}_{L} \W_{R}(x,0,\Lambda)=\P_L^{-1} \P_R(x,0,\Lambda).
\end{equation}
This implies
\begin{align}\label{moment identity}
&(\P_L^{-1} \P_R)(t)\\ \notag
&=
 \exp\left(
\sum_{n\geq 0} \sum_ {\alpha=1}^{N}
\Lambda^{N({n+1-\frac{\alpha-1}{N}})}t_{\alpha, n} \right)\circ(\P_L^{-1} \P_R)(0)\circ\exp\left(\sum_{n\geq 0} \sum_
{\beta=-M+1}^{0}\Lambda^{-M({n+1+\frac{\beta}{M}})}\,t_{\beta,
n}
\right).
\end{align}
If we let $\tau_0=1, \tau_i=0(i\in \Z_-);$ then bi-infinite matrix everywhere will become semi-infinite matrix.
Representing the product $\P_L^{-1}\P_R$ in the semi-infinite matrix,
 i.e. $\tilde\P_L^{-1}\tilde\P_R$ and considering identity \eqref{momentL} and identity \eqref{momentR}, eq.\eqref{moment identity} can be written as following
\begin{align}\label{moment}
&(\tilde\P_L^{-1} \tilde\P_R)(t)\\ \notag
&=\left(\begin{array}{ccccc}1 &P_1(t_{\alpha}) &P_2(t_{\alpha})  &\dots \\
0 &1  &P_1(t_{\alpha}) &\dots
\\ 0&0 &1  &\dots \\ \dots &\dots  &\dots&\dots
\end{array}\right) \left(\tilde\P_L^{-1} \tilde\P_R(0)\right)
\left(\begin{array}{ccccc}1 &0&0&\dots \\ P_1(t_{\beta})&1 &0&\dots
\\ P_2(t_{\beta})&P_1(t_{\beta}) &1 &\dots \\ \dots &\dots &\dots &\dots
\end{array}\right),
\end{align}
where $P_{k}$ are Schur polynomial as defined in \eqref{shurfunction}.
To construct tau functions, we define the {\it moment} matrix $M_{\infty}(t)$ as,
\begin{equation*}
M_{\infty}(t):=(\tilde\P_L^{-1} \tilde\P_R)(t).
\end{equation*}
\begin{proposition}
The constrained condition of BTH is equivalent to matrix $M_{\infty}$ satisfies identity
\begin{eqnarray}\label{moment matrix constraint}
\La^{Nn}M_{\infty}=M_{\infty}\La^{-Mn}.
\end{eqnarray}
\end{proposition}
\begin{proof}
Using constraint on $\L$, i.e. $\d_{t_{1,n}}\L=\d_{t_{0,n}}\L$, we can get
 $\d_{t_{1,n}}M_{\infty}=\d_{t_{0,n}}M_{\infty}$ which can lead to eq.\eqref{moment matrix constraint}.
\end{proof}
Direct calculation can lead to following proposition.
\begin{proposition}
The  matrix $M_{\infty}$ satisfies identity
\begin{eqnarray*}
\d_{t_{\alpha,n}} M_{\infty}&=&\Lambda^{N({n+1-\frac{\alpha-1}{N}})}M_{\infty}\\
\d_{t_{\beta,n}} M_{\infty}&=&M_{\infty}
\Lambda^{-M({n+1+\frac{\beta}{M}})}.
\end{eqnarray*}
\end{proposition}

Let $M_i$ be the $i\times i$ submatrix  of $M_{\infty}$ of the top left corner.
By eq.\eqref{moment}, each $\tau$-function can be obtained by the determinant \cite{ISo},
\begin{equation*}
\tau_i(t)=\det\left(M_i(t)\right).
\end{equation*}
Now, we will consider the detailed structure of tau functions of BTH from the point of reduction of the two-dimensional Toda hierarchy.

As we all know, the tau functions of the two-dimensional Toda lattice hierarchy are given by
\begin{equation}\label{tau n2Dtoda}
\tau_{i}=\left|\begin{array}{ccccc}\bar C_{0,0} & \bar C_{0,1}&\dots &\bar C_{0,i-1}\\ \bar C_{1,0} &\bar C_{1,1} &\dots &\bar C_{1,i-1}\\ \dots &\dots &\dots &\dots \\
\bar C_{i-1,0} &\bar C_{i-1,1}&\dots&\bar C_{i-1,i-1}\end{array}\right|,
\end{equation}
where
\begin{eqnarray*}
\bar C_{i,j}&=&\int\int\rho(\la,\mu)\la^{i}\mu^{j}e^{\sum_{n=0}^{\infty}x_n\la^n+\sum_{n=0}^{\infty}y_n\mu^n}d \la d\mu\\
&=&\sum_{k,l=0}^{\infty}\bar c_{i,j,k,l}P_k(x)P_l(y).
\end{eqnarray*}
We should note here that the coefficients $\bar c_{i,j,k,l}$ are totally independent.

As the original tridiagonal Toda lattice is $(1,1)$ reduction of the two-dimensional Toda lattice hierarchy. Therefore to get the solution of  tridiagonal Toda lattice, we need to add factor
$\delta(\la-\mu)$ under the integral in the definition of $\bar C_{i,j}$, i.e.
\begin{eqnarray}\label{11BTH}
\int\int\rho(\la,\mu)\delta(\la-\mu)\la^{i}\mu^{j}e^{\sum_{n=0}^{\infty}x_n\la^n+\sum_{n=0}^{\infty}y_n\mu^n}d \la d\mu,
\end{eqnarray}
which can further lead to
\begin{eqnarray}\label{11BTH2}
\int\rho(\la,\la)\la^{i+j}e^{\sum_{n=0}^{\infty}(x_n+y_n)\la^n}d \la.
\end{eqnarray}
After changing $x,y$ time variables to $t_{\alpha},t_{\beta}$, eq.\eqref{11BTH2} become a new function
$$\int\rho(\la,\la)\la^{i+j}e^{\xi_L(\la,t_{\alpha})+\xi_R(\la,t_{\beta})}d \la$$ which corresponds to  $(1,1)$-BTH.

Denote $\omega_N$ and $\omega_M$ as the N-th root and M-th root of unit.
For $(N,M)$-BTH, new function $C_{i,j}$(new form of $\bar C_{i,j}$) have the following form
\begin{eqnarray*}
C_{i,j}&=&\int\int\rho(\la,\mu)\delta(\la^N-\mu^M)\la^{i}\mu^{j}e^{\xi_L(\la,t_{\alpha})+\xi_R(\mu,t_{\beta})}d \la d\mu\\
&=&\sum_{p=0}^{N-1}\sum_{q=0}^{M-1}\int\rho(\omega_N^p\la^{\frac 1N},\omega_M^q\la^{\frac 1M})(\omega_N^p\la^{\frac 1N})^{i}(\omega_M^q\la^{\frac 1M})^{j}e^{\xi_L(\la^{\frac 1N}, t_{\alpha}^p)
+\xi_R(\la^{\frac 1M}, t_{\beta}^q)}d \la .
\end{eqnarray*}
where
\begin{eqnarray}\label{twist1}
 t_{\alpha,n}^p&=&(\omega_N^p)^{N(n+1-\frac {\alpha}N)}t_{\alpha,n}\\
 \label{twist2}
 t_{\beta,n}^q&=&(\omega_M^q)^{M(n+1+\frac {\beta}M)}t_{\beta,n}.
\end{eqnarray}
Transforms \eqref{twist1} and \eqref{twist2} can be seen as the twist of exponential solutions because of the bigraded structure.

 Because in the next we will consider rational solutions of BTH, we write  $C_{i,j}$ further into
\begin{eqnarray*}
C_{i,j}
=\sum_{p=0}^{N-1}\sum_{q=0}^{M-1}\sum_{k,l=0}^{\infty}c_{i,j,p,q}^{k,l}P_k(t_{\alpha})P_l(t_{\beta}),
\end{eqnarray*}
where
\begin{eqnarray*}
c_{i,j,p,q}^{k,l}&=&\int\rho(\omega_N^p\la^{\frac 1N},\omega_M^q\la^{\frac 1M})(\omega_N^p\la^{\frac 1N})^{i}(\omega_M^q\la^{\frac 1M})^{j}(\omega_N^p\la^{\frac 1N})^k(\omega_M^q\la^{\frac 1M})^ld \la.
\end{eqnarray*}
We can find coefficients $\{c_{i,j,p,q}^{k,l}; i,j,k,l\geq 0; 0\leq p\leq N-1, 0\leq q\leq M-1\}$ satisfy
\begin{eqnarray}\label{index relation'}
c_{i,j,p,q}^{k+N,l}&=&c_{i,j,p,q}^{k,l+M},
\end{eqnarray}
which tells us that $P_m(t_{\alpha})P_{n+M}(t_{\beta})$ and $P_{m+N}(t_{\alpha})P_{n}(t_{\beta})$ always appear at the same time.
So the element in position of $k$ row and $l$ column in initial moment matrix can have the following form
\begin{eqnarray}\label{index relation}
\left(\tilde\P_L^{-1} \tilde\P_R(0)\right)_{k,l}&=&\sum_{p=0}^{N-1}\sum_{q=0}^{M-1}c_{0,0,p,q}^{k-1,l-1}.
\end{eqnarray}

Therefore tau functions of the BTH  can be explicitly written in the form
\begin{equation}\label{tau n}
\tau_{i}=\left|\begin{array}{ccccc}C_{0,0} & C_{0,1}&\dots &C_{0,i-1}\\ C_{1,0} &C_{1,1} &\dots &C_{1,i-1}\\ \dots &\dots &\dots &\dots \\
C_{i-1,0} &C_{i-1,1}&\dots&C_{i-1,i-1}\end{array}\right|.
\end{equation}

With the definition of $C_{i,j}$, in the next part we will construct Lax matrix solution using orthogonal polynomials which are nothing but wave function in semi-infinite vector form. Before we do that, we need some formula about property of Schur Hirota derivatives  described in the following lemma \cite{Ohta}.
\begin{lemma}\label{young formula}
Schur derivatives have following formula
\begin{eqnarray}\label{D_L formular}
P_n({\hat D_L})\tau_i\cdot \tau_j=\sum_{k+l=n}P_k(\hat\partial_L)\tau_i\times P_l(-\hat\partial_L)\tau_j,
\end{eqnarray}
\begin{eqnarray}
P_l(\hat\partial_L)[0,1,2,\ldots, i-1]_L & =&  [0,1,2,..., i-2, i+l-1]_L,\\
P_l(-\hat\partial_L) [0,1,2,\ldots,i-1]_L& = &(-1)^l [0,1,...., i-l-1, i-l+1,....,i-1,i]_L,
\end{eqnarray}
\begin{eqnarray}\label{D_R formular}
P_n({\hat D_R})\tau_i\cdot \tau_j=\sum_{k+l=n}P_k(\hat\partial_R)\tau_i\times P_l(-\hat\partial_R)\tau_j,
\end{eqnarray}
\begin{eqnarray}
P_l(\hat\partial_R)[0,1,2,\ldots, i-1]_R & =  &[0,1,2,..., i-2, i+l-1]_R,\\
P_l(-\hat\partial_R) [0,1,2,\ldots,i-1]_R& =& (-1)^l [0,1,...., i-l-1, i-l+1,....,i-1,i]_R,
\end{eqnarray}
where
\begin{equation}
[k_0,k_1,k_2,\ldots, k_{i-1}]_L = \left|\begin{array}{ccccc}C_{k_0,0} & C_{k_0,1}&\dots &C_{k_0,i-1}\\ C_{k_1,0} &C_{k_1,1} &\dots &C_{k_1,i-1}\\ \dots &\dots &\dots &\dots \\
C_{k_{i-1},0} &C_{k_{i-1},1}&\dots&C_{k_{i-1},i-1}\end{array}\right|,
\end{equation}
\begin{equation}
[k_0,k_1,k_2,\ldots, k_{i-1}]_R = \left|\begin{array}{ccccc}C_{0,k_0} & C_{0,k_1}&\dots &C_{0,k_{i-1}}\\ C_{1,k_0} &C_{1,k_1} &\dots &C_{1,k_{i-1}}\\ \dots &\dots &\dots &\dots \\
C_{i-1,k_0} &C_{i-1,k_1}&\dots&C_{i-1,k_{i-1}}\end{array}\right|,
\end{equation}
where $P_{k}$ are Schur polynomial as defined in \eqref{shurfunction}.
\end{lemma}

In fact, Lemma \ref{young formula} is a special case of abstract general formula of Schur function \cite{Ohta}
\begin{eqnarray}\label{tauphi}
S_Y(\hat \d)\tau_\phi=\tau_Y(t).
\end{eqnarray}
Here $\tau_\phi:=[0,1,2,\ldots, i-1]$ is the standard Wronskian determinant,  $Y:=(Y_0,Y_1,Y_2,\ldots, Y_{i-1})$  and
 \begin{eqnarray*}
S_{Y}&=&\left|\begin{array}{ccccc}  P_{Y_{i-1}+i-1} &  P_{Y_{i-2}+i-2}&\dots & P_{Y_1+1}&  P_{Y_0}\\  P_{Y_{i-1}+i-2}&  P_{Y_{i-2}+i-3}&\dots & P_{Y_1} & P_{Y_0-1}\\ \dots&\dots &\dots &\dots &\dots \\
P_{Y_{i-1}+1} &   P_{Y_{i-2}}&\dots & P_{Y_1-i+3}&  P_{Y_0-i+2}\\
 P_{Y_{i-1}} &   P_{Y_{i-2}-1}&\dots & P_{Y_1-i+2}&  P_{Y_0-i+1}\end{array}\right|_{j\times j}.
\end{eqnarray*}
 $Y=(Y_0,Y_1,Y_2,\ldots, Y_{i-1})\ (Y_0>Y_1>Y_2>\ldots> Y_{i-1})$ corresponds to Young diagram with $Y_0$ boxes at the first row, $Y_1$ boxes at the second row and so on.
Notice formula \cite{Ohta}
\begin{eqnarray} \label{tau diagram}
S_Y(-t)=(-1)^{|Y|}S_{Y'}(t),
\end{eqnarray}
where $Y'$ is the conjugate Young diagram of  $Y$.
Eq.\eqref{tauphi} together with eq.\eqref{tau diagram} further leads to the lemma above easily.

Then we define  wave functions $W_L=(W_{L1},W_{L2},\ldots)$, $\bar W_{R}=(\bar W_{R1},\bar W_{R2},\ldots)^T$ and
$\hat W_{R}=(\hat W_{R1},\hat W_{R2},\ldots)^T$ with
 \begin{align}\label{W_{Li}}
W_{Li}(\la^{\frac1N},t)&= \frac{ e^{\xi_L(\la^{\frac1N},t)}}{\tau_{i-1}}\left|\begin{array}{ccccc}C_{0,0} & C_{0,1}&\dots &C_{0,i-2}&1\\ C_{1,0} &C_{1,1} &\dots &C_{1,i-2}&\la^{\frac1N}\\ \dots &\dots &\dots &\dots &\dots\\
C_{i-1,0} &C_{i-1,1}&\dots&C_{i-1,i-2}&\la^{\frac {i-1}N}\end{array}\right|,
\end{align}
\begin{align}\label{W_{Ri}}
\bar W_{Rj}(\la^{\frac1M},t)&= \frac{e^{\xi_R(\la^{\frac1M},t)}}{\tau_j}
\left|\begin{array}{cccc}C_{0,0} & C_{0,1}&\dots&  C_{0,j-1} \\ C_{1,0} &C_{1,1}&\dots & C_{1,j-1} \\\dots &\dots &\dots &\dots\\
 C_{j-2,0} &C_{j-2,1}&\dots & C_{j-2,j-1} \\1 & \la^{\frac1M}&\dots & \la^{\frac {j-1}M}\end{array}\right|,
 \end{align}
\begin{align}
\hat W_{Rj}(\la^{\frac1M},t)&= \frac{e^{\xi_R(\la^{\frac1M})}}{\tau_{j-1}}
\left|\begin{array}{cccc}C_{0,0} & C_{0,1}&\dots&  C_{0,j-1} \\ C_{1,0} &C_{1,1}&\dots & C_{1,j-1} \\\dots &\dots &\dots &\dots\\
 C_{j-2,0} &C_{j-2,1}&\dots & C_{j-2,j-1} \\1 & \la^{\frac1M}&\dots & \la^{\frac {j-1}M}\end{array}\right|,
\end{align}
which satisfy the following orthogonality relation,
\begin{eqnarray}
\langle W_{Li}, \bar W_{Rj}\rangle=\delta_{i,j},\ \ \ \langle W_{Li}, \hat W_{Rj}\rangle=\delta_{i,j}h_j,
\end{eqnarray}
where
\begin{eqnarray*}h_j:=\frac{\tau_j}{\tau_{j-1}},
\end{eqnarray*}
and inner product $\langle ,  \rangle$ of functions $A$ and $B$ is defined as
\begin{eqnarray*}
\langle A, B \rangle:=\int\int\rho(\la,\mu)\delta(\la^N-\mu^M)A(\la,t)B(\mu,t)d \la d\mu.
\end{eqnarray*}
Therefore tau functions have another form as
\begin{eqnarray*}\tau_m:=det\left(\langle W_{Li}, \hat W_{Rj}\rangle\right)_{1\leq i,j\leq m}.
\end{eqnarray*}
The entries of the matrix representation of the Lax operator $\L$ can be then calculated by
\begin{eqnarray*}
a_{ij}&=&\frac{P_{i-j+N}(\hat D_L)\tau_j\tau_{i-1}}{\tau_{i-1}\tau_j}
\\&=&\frac{1}{\tau_{i-1}\tau_j}\sum_{m+n=i-j+N}P_m(\hat \d_L)\tau_jP_n(-\hat \d_L)\tau_{i-1}\\
&=&\frac{1}{\tau_{i-1}\tau_j}\sum_{m=0}^{i-j+N}\[0,1,\dots, j-2, j+m-1\]_L\\
&&(-1)^{m+j-i-N}\[0,1,\dots j-N+m-2, j-N+m,\dots ,i-1\]_L\\
&=&\langle\frac{\la e^{\xi_L(\la^{\frac1N},t)}}{\tau_{i-1}}\left|\begin{array}{ccccc}C_{0,0} & C_{0,1}&\dots &C_{0,i-2}&1\\ C_{1,0} &C_{1,1} &\dots &C_{1,i-2}&\la^{\frac1N}\\ \dots &\dots &\dots &\dots &\dots\\
C_{i-1,0} &C_{i-1,1}&\dots&C_{i-1,i-2}&\la^{\frac {i-1}N}\end{array}\right|, \frac{e^{\xi_R(\la^{\frac1M},t)}}{\tau_j}
\left|\begin{array}{cccc}C_{0,0} & C_{0,1}&\dots&  C_{0,j-1} \\ C_{1,0} &C_{1,1}&\dots & C_{1,j-1} \\\dots &\dots &\dots &\dots\\
 C_{j-2,0} &C_{j-2,1}&\dots & C_{j-2,j-1} \\1 & \la^{\frac1M}&\dots & \la^{\frac {j-1}M}\end{array}\right|\rangle.
\end{eqnarray*}
So
\begin{equation}\label{aijformula}
a_{i,j}=\langle\la W_{Li}, \bar W_{Rj}\rangle
\end{equation}
which are given by  the matrix representations of the eigenvalue problems $\L W_L=\lambda W_L$ and $ \bar W_R\L=\lambda \bar W_R$(see \cite{ISo} for the details). Till now, we have solved the  BTH using orthogonal polynomials.

If we denote $\tilde \W_L$ and $\tilde\W_R^{-1}$ as corresponding matrix forms of  $\W_L$ \eqref{W_L} and $\W_R^{-1}$ \eqref{W_R} respectively,
then  $W_L$ and $\bar W_R$ can be represented by matrices $\tilde\W_L$ and $\tilde\W_R^{-1}$ respectively as following.

Because
\begin{eqnarray*}
\La\left(\begin{array}{cccc}1  \\ \la^{\frac1N} \\\la^{\frac2N}\\
 \cdot \\ \cdot \end{array}\right)&=&\la^{\frac1N}\left(\begin{array}{cccc}1  \\ \la^{\frac1N} \\\la^{\frac2N}\\
 \cdot \\ \cdot \end{array}\right),
  \end{eqnarray*}
 we can get
\begin{eqnarray*}
W_L&=&\tilde\W_L\left(\begin{array}{cccc}1  \\ \la^{\frac1N} \\\la^{\frac2N}\\
 \cdot \\ \cdot \end{array}\right)
 =\tilde\P_L(x,t,\Lambda)
 \left(\begin{array}{cccc}1  \\ \la^{\frac1N} \\\la^{\frac2N}\\
 \cdot \\ \cdot \end{array}\right)e^{\xi_L(\la^{\frac1N},t)}\\
 &=&
 \left(\begin{array}{cccc}1  \\ \frac{P_1(-\hat \d_L)\tau_1}{\tau_1} +\la^{\frac1N} \\\frac{P_2(-\hat \d_L)\tau_2}{\tau_2}+
 \frac{P_1(-\hat \d_L)\tau_2}{\tau_2}\la^{\frac1N}+\la^{\frac2N}\\
 \cdot \\ \cdot \end{array}\right)e^{\xi_L(\la^{\frac1N},t)};
  \end{eqnarray*}
  where $\tilde\P_L(x,t,\Lambda)$ is as matrix \eqref{momentL1}.
 This agree with the definition of $W_{Li}$ in eq.\eqref{W_{Li}}.
  Also similarly we can get
  \begin{eqnarray*}
 \bar W_R&=&\left(\begin{array}{ccccc}1 & \la^{\frac1M} &\la^{\frac2M}&
 \cdot & \cdot \end{array}\right)\tilde\W_R^{-1}=e^{\xi_R(\la^{\frac1M},t)}\left(\begin{array}{ccccc}1 & \la^{\frac1M} &\la^{\frac2M}&
 \cdot & \cdot \end{array}\right)\tilde\P_R^{-1}(x,t,\Lambda)\\
 &=&e^{\xi_R(\la^{\frac1M},t)}\left(\begin{array}{ccccc}\frac{\tau_0}{\tau_1}, &\frac{P_1(-\hat \d_R)\tau_1}{\tau_2} + \la^{\frac1M}\frac{\tau_1}{\tau_2}, &\frac{P_2(-\hat \d_R)\tau_2}{\tau_3}+\frac{P_1(-\hat \d_R)\tau_2}{\tau_3} \la^{\frac1M}+
 \frac{\tau_2}{\tau_3}\la^{\frac2M},&
 \cdot & \cdot \end{array}\right),
 \end{eqnarray*}
   where $\tilde\P_R^{-1}(x,t,\Lambda)$ is as matrix \eqref{momentR2}.
This also agrees with the definition of $\bar W_{Ri}$ in eq.\eqref{W_{Ri}}.

 Formal factorization mentioned above is  about infinite-sized Lax matrix. In the next section, we will consider its finite-sized truncation. Then we can find finite-sized Lax matrix is in fact nilpotent because of dressing structure \eqref{two dressing}. This finite-sized case corresponds to rational solutions of the BTH which will be considered in the next section.

\sectionnew{Rational solutions of the $(N,M)$-BTH}\label{sec:Rational}

It is well known that the $\tau$-function of the original tridiagonal Toda lattice, i.e. $(1,1)$-BTH,
has the Schur polynomial solutions associated with rectangular Young diagrams. So what kind of Young diagrams correspond to
the BTH become an interesting question. In this section, we only consider homogeneous rational solution of $(N,M)$-BTH
 which is one kind  of  most interesting solutions in nonlinear integrable systems.

In order to describe homogeneous rational solution of $(N,M)$-BTH $(N\leq M)$, we firstly set
$deg(t_{1,0})=deg(t_{0,0})=MN.$ Then we get $deg(P_m(t_{\alpha}))= mM$ and $deg(P_n(t_{\beta}))= nN$.\\
Define
 \begin{eqnarray*}c_{p,q}^{m,n}:=c_{0,0,p,q}^{m,n},
\end{eqnarray*}
and  choose the following special homogeneous polynomials $\bar P_k(t_{\alpha},t_{\beta})$ with degree $k$ as $\tau_1$,
 \begin{equation*}
\bar P_k(t_{\alpha},t_{\beta}):=\sum_{p=0}^{N-1}\sum_{q=0}^{M-1}\sum_{mM+nN=k}c_{p,q}^{m,n}P_m(t_{\alpha})P_n(t_{\beta}),
\end{equation*}
where $t_{\alpha}$ and   $t_{\beta}$ denote
\begin{align*}
t_{\alpha}&=\{t_{\alpha,n}:1\le\alpha\le N, n=0,1,2,\ldots\}\\
t_{\beta}&=\{t_{\beta,n}: -M+1\le\beta\le 0, n=0,1,2,\ldots\},
\end{align*}
and  $k$ can be any number in the set $\{mM+nN; m,n\in\Z_+\}.$
The polynomials $P_m(t_{\alpha})$ and $P_n(t_{\beta})$ are the elementary Schur polynomials, and
they satisfy the following relations,
\begin{equation*}
\frac{\partial P_m(t_{\alpha})}{\partial t_{\alpha',m'}}=P_{m-N(m'+1)+\alpha'-1}(t_{\alpha}),\qquad
\frac{\partial P_n(t_{\beta})}{\partial t_{\beta',n'}}=P_{n-M(n'+1)-\beta'}(t_{\beta}).
\end{equation*}
Note here that $\{c_{p,q}^{m,n}|0\leq m,p\leq N-1, 0\leq n,q\leq M-1\}$ can be  arbitrary constants.\\

Define
\begin{eqnarray*}\bar P_k^{l,l'}=\sum_{p=0}^{N-1}\sum_{q=0}^{M-1}\sum_{mM+nN=k}c_{p,q}^{m+l,n+l'}P_m(t_{\alpha})P_n(t_{\beta}),
\end{eqnarray*}
and the rational solutions for (N,M)-BTH have the following  diagram representation,
\begin{eqnarray*}D_j=\{k-(j-1)N,k-(j-2)N-M,\dots,k-(j-1)M\}, k=0,N,M,2N,N+M,2M,\dots.
\end{eqnarray*}
The difference between two adjacent two numbers in $D_j$ is $M-N$.

We denote the tau function corresponding to $D_j(k)$ as $\tau_{j,D_j}$ which have the following form
\begin{eqnarray*}
\tau_{j,D_j(k)}=S_{D_j(k)}&=&\left|\begin{array}{ccccc}\bar P_k^{(0,0)} & \bar P_{k-M}^{(1,0)}&\dots &\bar P_{k-(j-1)M}^{(j-1,0)}\\ \bar P_{k-N}^{(0,1) }& \bar P_{k-M-N}^{(1,1)}&\dots &\bar P_{k-(j-1)M-N}^{(j-1,1)}\\ \dots &\dots &\dots &\dots \\
\bar P_{k-(j-1)N}^{(0,j-1)} & \bar P_{k-M-(j-1)N}^{(1,j-1)}&\dots &\bar P_{k-(j-1)(M+N)}^{(j-1,j-1)}\end{array}\right|_{j\times j},
\end{eqnarray*}
where $\tau_{j,D_j(k)}$ denotes the j-th tau function($j\times j$ determinant) generated by $\bar P_k^{(0,0)}$.
The range of rank $j$ depends on the choice of $k$.
This kind of diagram like $D_j$  is not classical Young diagram. It is a kind of generalized  diagram which counts the homogeneous degree which  comes from the multiplication of two classical Shur functions. We can call this  kind of generalized  diagram {\emph{degree diagram}}. Same as Young diagram, the tau functions represented by degree diagram  is also Wronskian form, the derivative is about $\d_{t_{-M+1,0}}$ or $\d_{t_{N,0}}$. Because the scale of degree in definition before(e.g. $deg(P_m(t_{\alpha}))= mM$) is bigger than common degree of Schur polynomial($deg(P_m(t))= m$), the difference of subscript between two adjacent
 rows(columns) is $N(M)$ not $1(1)$. From this point, it is also different from Hankel determinant.

In the following we firstly only  consider the case when $N$ and $M$ are co-prime integers. When they are not co-prime, just divide them by the GCD  of them and use the theory of co-prime case in the following.
In fact, we can find for fixed value of $p,q$, all the coefficients of  a homogeneous polynomial will be the same because they all satisfy relation \eqref{index relation'}. Then we get
\begin{eqnarray*}
\bar P_k^{l,l'}&=&\sum_{p=0}^{N-1}\sum_{q=0}^{M-1}\sum_{mM+nN=k}c_{p,q}^{m+l,n+l'}P_m(t_{\alpha})P_n(t_{\beta})\\
&=&\sum_{p=0}^{N-1}\sum_{q=0}^{M-1}c_{p,q}^{m_1+l,n_1+l'}P_{m_1}(t_{\alpha})P_{n_1}(t_{\beta})+c_{p,q}^{m_2+l,n_2+l'}P_{m_2}(t_{\alpha})
P_{n_2}(t_{\beta})+\dots\\
&&+c_{p,q}^{m_{l(k)}+l,n_{l(k)}+l'}P_{m_{l(k)}}(t_{\alpha})P_{n_{l(k)}}(t_{\beta})\\
&=&\sum_{p=0}^{N-1}\sum_{q=0}^{M-1}c_{p,q}^{m_1+l,n_1+l'}\sum_{mM+nN=k}P_m(t_{\alpha})P_n(t_{\beta})\\
&=&c^{l,l'}_k\sum_{mM+nN=k}P_m(t_{\alpha})P_n(t_{\beta}),
\end{eqnarray*}
where
\begin{eqnarray*}
c_{p,q}^{m_1+l,n_1+l'}&=&c_{p,q}^{m_2+l,n_2+l'}=\dots=c_{p,q}^{m_{l(k)}+l,n_{l(k)}+l'},\\
c^{l,l'}_k&=&\sum_{p=0}^{N-1}\sum_{q=0}^{M-1}c_{p,q}^{m_1+l,n_1+l'},
\end{eqnarray*}
\begin{eqnarray*}
m_iM+n_iN=k,\ \
m_{i+1}=m_i-N,\ \ n_{i+1}=n_i+M,\ \ 1\leq i\leq l(k),
\end{eqnarray*}
and number $l(k)$ depends on $k$.
For simplicity, here we just consider the case with all coefficients $c^{l,l'}_k$ equal $1$ which is our central consideration in this section and define
\begin{eqnarray*}\bar P_k=\sum_{mM+nN=k}P_m(t_{\alpha})P_n(t_{\beta})=
P_{m_1}(t_{\alpha})P_{n_1}(t_{\beta})+P_{m_2}(t_{\alpha})P_{n_2}(t_{\beta})+\dots+P_{m_{l(k)}}(t_{\alpha})P_{n_{l(k)}}(t_{\beta}).
\end{eqnarray*}

Then the $\tau$-function $\tau_s$ generated by $\tau_1=\bar P_k $ can be expressed by a double-Wronskian determinant,
\begin{eqnarray*}\label{tauRational}
\tau_s(k,t_{\alpha},t_{\beta})&=&\left|\begin{array}{ccccc}\bar P_k & \bar P_{k-M}&\dots &\bar P_{k-(s-1)M}\\ \bar P_{k-N}& \bar P_{k-M-N}&\dots &\bar P_{k-(s-1)M-N}\\ \dots &\dots &\dots &\dots \\
\bar P_{k-(s-1)N} & \bar P_{k-M-(s-1)N}&\dots &\bar P_{k-(s-1)(M+N)}\end{array}\right|_{s\times s}\\
&=&\left|\left(\begin{array}{ccccc} P_{m_1}(t_{\alpha}) & P_{m_2}(t_{\alpha})&\dots &P_{m_l}(t_{\alpha})\\ P_{m_1-1}(t_{\alpha})& P_{m_2-1}(t_{\alpha})&\dots &P_{m_l-1}(t_{\alpha})
\\ P_{m_1-2}(t_{\alpha})& P_{m_2-2}(t_{\alpha})&\dots &P_{m_l-2}(t_{\alpha})\\ \dots &\dots &\dots &\dots \\
P_{m_1-s+1}(t_{\alpha}) & P_{m_2-s+1}(t_{\alpha})&\dots & P_{m_l-s+1}(t_{\alpha})\end{array}\right)\right.\\
&&\left.\times
\left(\begin{array}{ccccc} P_{n_1}(t_{\beta}) & P_{n_1-1}(t_{\beta})& P_{n_1-2}(t_{\beta})&\dots &P_{n_1-s+1}(t_{\beta})\\ P_{n_2}(t_{\beta}) & P_{n_2-1}(t_{\beta})& P_{n_2-2}(t_{\beta})&\dots &P_{n_2-s+1}(t_{\beta})
\\ \dots &\dots &\dots &\dots \\
P_{n_l}(t_{\beta}) & P_{n_l-1}(t_{\beta})& P_{n_l-2}(t_{\beta})&\dots &P_{n_l-s+1}(t_{\beta})\end{array}\right)\right|_{s\times s}.
\end{eqnarray*}

 Conversely, for a fixed size $j$ of Lax matrix for $(N,M)$-BTH, the choices of $k$ for $\tau_1$ is in set $K_j$
\begin{equation}\label{K_j}
K_j:=\left\{k| k =(j-1)NM+mM+nN, m,n\in \Z_+, 0\leq m< N, 0\leq n<M \right\}.
\end{equation}
We can see that the number of elements in set $K_j$ is $NM$(When they are not co-prime, this number will be $\frac{NM}{(N,M)^2}$, where $(N,M)$ is GCD of $N$ and $M$).
When the values  of $N,M,j,m,n$ are chosen, a series of non-vanishing tau functions corresponding  to them will be fixed.

 In the following, we will consider rational solutions of $(N,M)$-BTH with finite-sized Lax matrix.
For $(N,M)$-BTH, the size of minimal Lax matrix is $(M+1)\times(M+1).$
This minimal Lax matrix has $M$ non-vanishing $\tau$-functions and anyone's degree has $M-N$ jumps between adjacent rows. For the special case of $N=M$, degree diagrams are always rectangle which can also be seen from definition of $D_j$.

Besides considering the degree diagrams, it is more interesting to consider the decomposition of degree diagrams into representation of
Young diagrams. In fact the Young diagram representation of general BTH has a form of multiplication of two different groups of Young diagrams which will be shown in the following theorem.
\begin{theorem}
For $j\times j (j\geq M+1)$-sized Lax matrix of  $(N,M)$-BTH (denoted as $(N,M)_{j\times j}$), after choosing the value of $k$ as $(j-1)MN+mM+nN$, Young diagram representation of a series of corresponding tau functions are  as following
\begin{eqnarray*}
\tau_1&=&\sum_{0\leq a\leq j-1}S_{((j-1-a)N+m)}(t_{\alpha})S_{(n+aM)}(t_{\beta}),\\
\tau_2&=&\sum_{ 0\leq a<b\leq j-1}S_{((j-1-a)N+m-1,(j-1-b)N+m)}(t_{\alpha})S_{(n+bM-1,n+aM)}(t_{\beta}),\\
\tau_3&=&\sum_{ 0\leq a<b<c\leq j-1}S_{((j-1-a)N+m-2,(j-1-b)N+m-1,(j-1-c)N+m)}(t_{\alpha})S_{(n+cM-2,n+bM-1,n+aM)}(t_{\beta}),\\ \notag
\dots&&\dots \dots\dots\\
\tau_s&=&\sum_{ 0\leq a_1<a_2<\dots<a_s\leq j-1}S_{((j-1-a_1)N+m-s+1,(j-1-a_2)N+m-s+2,\dots,(j-1-a_s)N+m)}(t_{\alpha})\\
&&S_{(n+a_sM-s+1,\dots,n+a_2M-1,n+a_1M)}(t_{\beta}),\\ \notag
\dots&&\dots \dots\dots\\
\tau_j&=&S_{((j-1)(N-1)+m,(j-2)(N-1)+m,\dots,m)}(t_{\alpha})S_{(n+(j-1)(M-1),n+(j-2)(M-1),\dots,n)}(t_{\beta}).
\end{eqnarray*}
\end{theorem}
\begin{proof}
To prove this theorem, one need to use Cauchy-Binet formula. The process is quite complicated because of huge sizes of matrices. So we will omit the proof. One can understand the patten by the following example, i.e. $(2,3)$-BTH.
\end{proof}
 To see it clearly, we give some specific examples in the following.
\begin{Example}
$(2,3)$-BTH,   has 6 sets of $\tau$-functions for each size of Lax matrix and the degree diagram for every tau function has one jump between
adjacent rows. See $(2,3)_{4\times 4}$ in detail as following degree diagram

\begin{equation}\label{23,4}
(2,3)_{4\times 4}
\begin{cases}
(0,0)&:\ \  \tau_{1,\{18\}}\rightarrow \ \ \tau_{2,\{16,15\}}\rightarrow \ \ \tau_{3,\{14,13,12\}}\rightarrow \ \ \tau_{4,\{12,11,10,9\}},\\
(0,1)&:\ \  \tau_{1,\{20\}}\rightarrow \ \ \tau_{2,\{18,17\}}\rightarrow \ \ \tau_{3,\{16,15,14\}}\rightarrow \ \ \tau_{4,\{14,13,12,11\}},\\
(1,0)&:\ \  \tau_{1,\{21\}}\rightarrow \ \ \tau_{2,\{19,18\}}\rightarrow \ \ \tau_{3,\{17,16,15\}}\rightarrow \ \ \tau_{4,\{15,14,13,12\}},\\
(0,2)&:\ \  \tau_{1,\{22\}}\rightarrow \ \ \tau_{2,\{20,19\}}\rightarrow \ \ \tau_{3,\{18,17,16\}}\rightarrow \ \ \tau_{4,\{16,15,14,13\}},\\
(1,1)&:\ \  \tau_{1,\{23\}}\rightarrow \ \ \tau_{2,\{21,20\}}\rightarrow \ \ \tau_{3,\{19,18,17\}}\rightarrow \ \ \tau_{4,\{17,16,15,14\}},\\
(1,2)&:\ \  \tau_{1,\{25\}}\rightarrow \ \ \tau_{2,\{23,22\}}\rightarrow \ \ \tau_{3,\{21,20,19\}}\rightarrow \ \ \tau_{4,\{19,18,17,16\}},\\
\end{cases}
\end{equation}
where $\{(p,q),0\leq p<2,0\leq q<3\}$ denote the value of $(m,n)$ in value of $k$, i.e. \eqref{K_j}. Here $k$ takes values in $\{18,20,21,22,23,25\}$, i.e. the values in bracket of $\tau_{1,\{\}}$. Every tau function $\tau_{1,\{\}}$ generates a series of tau functions which are connected by right arrow in \eqref{23,4}. $\tau_{l,\{\ ,\dots, \  \}}$ represents the l-th tau function whose degree diagram is in the bracket $\{\ ,\dots, \  \}$.
In  \eqref{23,4}, we can use product of Young diagrams to represent the four tau functions of the first line, i.e. $(0,0)$ case as following  by Cauchy-Binet formula
\begin{eqnarray*}\label{23,4'}
 \tau_1&=&\tau_{1,\{18\}}=\left|\left(\begin{array}{ccccc} P_6(t_{\alpha}) & P_4(t_{\alpha})&P_2(t_{\alpha}) &1\end{array}\right)
 \left(\begin{array}{ccccc}  1 \\  P_3(t_{\beta}) \\P_6(t_{\beta}) \\
P_{9}(t_{\beta}) \end{array}\right)\right|\\
&=&S_{\mbox{\young[6][4][ , , , ]}}(t_{\alpha})S_{\phi}(t_{\beta})+S_{\mbox{\young[4][4][ , , , ]}}(t_{\alpha})S_{\young[3][4][ , , , ]}(t_{\beta})+S_{\mbox{\young[2][4][ , , , ]}}(t_{\alpha})S_{\young[6][4][ , , , ]}(t_{\beta})+S_{\phi}(t_{\alpha})S_{\young[9][4][ , , , ]}(t_{\beta});
\end{eqnarray*}
\begin{eqnarray*}
\tau_2&=&\tau_{2,\{16,15\}}=\left|\left(\begin{array}{ccccc} P_6(t_{\alpha}) & P_4(t_{\alpha})&P_2(t_{\alpha}) &1\\
 P_5(t_{\alpha}) & P_3(t_{\alpha})&P_1(t_{\alpha}) &0\end{array}\right)
 \left(\begin{array}{ccccc}  1 & 0\\  P_3(t_{\beta})& P_2(t_{\beta}) \\P_6(t_{\beta})&P_5(t_{\beta}) \\
P_{9}(t_{\beta})& P_8(t_{\beta}) \end{array}\right)\right|\\
&=&S_{\mbox{\young[5,4][4][ , , , ]}}(t_{\alpha})S_{\young[2][4][ , , , ]}(t_{\beta})+S_{\mbox{\young[5,2][4][ , , , ]}}(t_{\alpha})S_{\young[5][4][ , , , ]}(t_{\beta})+S_{\mbox{\young[5][4][ , , , ]}}(t_{\alpha})S_{\young[8][4][ , , , ]}(t_{\beta})\\
&&+S_{\mbox{\young[3,2][4][ , , , ]}}(t_{\alpha})S_{\young[5,3][4][ , , , ]}(t_{\beta})+S_{\mbox{\young[3][4][ , , , ]}}(t_{\alpha})S_{\young[8,3][4][ , , , ]}(t_{\beta})
+S_{\mbox{\young[1][4][ , , , ]}}(t_{\alpha})S_{\young[8,6][4][ , , , ]}(t_{\beta});
\end{eqnarray*}
\begin{eqnarray*}
\tau_3&=&\tau_{3,\{14,13,12\}}=\left|\left(\begin{array}{ccccc} P_6(t_{\alpha}) & P_4(t_{\alpha})&P_2(t_{\alpha}) &1\\  P_5(t_{\alpha}) & P_3(t_{\alpha})&P_1(t_{\alpha}) &0
\\  P_4(t_{\alpha}) & P_2(t_{\alpha})&1 &0\end{array}\right)
 \left(\begin{array}{ccccc}  1 & 0&0\\  P_3(t_{\beta}) & P_2(t_{\beta})&P_1(t_{\beta})
\\P_6(t_{\beta}) & P_5(t_{\beta})&P_4(t_{\beta})  \\
P_{9}(t_{\beta}) & P_8(t_{\beta})&P_7(t_{\beta})\end{array}\right)\right|\\
&=&S_{\mbox{\young[4,3,2][4][ , , , ]}}(t_{\alpha})S_{\young[4,2][4][ , , , ]}(t_{\beta})+S_{\mbox{\young[4,3][4][ , , , ]}}(t_{\alpha})S_{\young[7,2][4][ , , , ]}(t_{\beta})+S_{\mbox{\young[4,1][4][ , , , ]}}(t_{\alpha})S_{\young[7,5][4][ , , , ]}(t_{\beta})\\
&&+S_{\mbox{\young[2,1][4][ , , , ]}}(t_{\alpha})S_{\young[7,5,3][4][ , , , ]}(t_{\beta});
\end{eqnarray*}
\begin{eqnarray*}
\tau_4&=&\tau_{4,\{12,11,10,9\}}=\left|\left(\begin{array}{ccccc} P_6(t_{\alpha}) & P_4(t_{\alpha})&P_2(t_{\alpha}) &1\\  P_5(t_{\alpha}) & P_3(t_{\alpha})&P_1(t_{\alpha}) &0
\\  P_4(t_{\alpha}) & P_2(t_{\alpha})&1 &0\\
 P_3(t_{\alpha}) & P_1(t_{\alpha})&0&0\end{array}\right)
 \left(\begin{array}{ccccc}  1 & 0&0&0\\  P_3(t_{\beta}) & P_2(t_{\beta})&P_1(t_{\beta}) &1
\\P_6(t_{\beta}) & P_5(t_{\beta})&P_4(t_{\beta}) &P_3(t_{\beta}) \\
P_{9}(t_{\beta}) & P_8(t_{\beta})&P_7(t_{\beta}) &P_6(t_{\beta})\end{array}\right)\right|\\
&=&S_{\mbox{\young[3,2,1][4][ , , , ]}}(t_{\alpha})S_{\young[6,4,2][4][ , , , ]}(t_{\beta}).
\end{eqnarray*}
\end{Example}

After general theory on homogeneous rational solutions of the $(N,M)$-BTH,
as a special but important case, rational solutions of the $(1,M)$-BTH will be considered in the next subsection.

\subsection{Rational solutions of the $(1,M)$-BTH}
Since  the $t_{1,n}$ flows are same as the $t_{0,n}$ flows, we use $t_{1,n}+t_{0,n}$ as a new variable and identify $t_{1,n}$ as $t_{0,n}$.
Then the rational solutions for the $(1,M)$-BTH are obtained from the $\tau$-function,
\begin{equation}
\tau_j(k,t)=\left|\begin{array}{ccccc}P_k & P_{k-M}&\dots &P_{k-(j-1)M}\\ P_{k-1} & P_{k-M-1}&\dots &P_{k-(j-1)M-1}\\ \dots &\dots &\dots &\dots \\
P_{k-(j-1)} & P_{k-M-(j-1)}&\dots &P_{k-(j-1)(M+1)}\end{array}\right|_{j\times j},
\end{equation}
where $P_n=P_n(t_{\beta})$ with $t_{0,n}\equiv t_{0,n}+t_{1,n}$.
This $\tau$-function can be given by the Schur polynomial associated with the Young diagram which is same as degree diagram mentioned above for
$(1,M)$-BTH, i.e.
$\tau_j(k)=S_{Y_j}(k)$ with
\begin{equation*}
Y_j(k)=(k-j+1, k-j+1-(M-1),\ldots,k-(j-1)M) \qquad {\rm for} \quad j=1,2,\ldots,1+\left\lfloor\frac{k}{M}\right\rfloor,
\end{equation*}
where $\left\lfloor\frac{k}{M}\right\rfloor$ denotes the biggest integer which is less than or equal $\frac{k}{M}$.
Note here that the number of boxes in the diagram increases by $M-1$ between adjacent rows.
Let us now give some examples
of the $\tau$-functions for the $(1,M)$-BTH with specific size $r$ of the Lax matrix, denoted by
$(1,M)_{r\times r}$.  For a given size $r(>M)$, the choices of $k$ are in the set $\{(r-1)M,\ldots, rM-1\}$, that is, there are $M$ choices of the first member of the $\tau$-functions, $\tau_1=P_k$.    Then each value of $k$  generates $r$ tau functions ordered from $\tau_1$ to $\tau_r.$
\begin{Example}
$(1,1)$-BTH, i.e. the original Toda lattice,  has only one set of $\tau$-functions for each size of Lax
matrix;
\begin{equation}\label{yp}
\begin{cases}
(1,1)_{2\times 2}&:\ \ \  \tau_{\mbox{\young[1][4][ , , , ]}}\rightarrow \ \ \tau_{\phi},\\
(1,1)_{3\times 3}&:\ \ \ \tau_{\mbox{\young[2][4][ , , , ]}}\rightarrow \ \ \tau_{\mbox{\young[1,1][4][ , , , ]}}\rightarrow \ \ \tau_{\phi},\\
(1,1)_{4\times 4}&:\ \ \tau_{\mbox{\young[3][4][ , , , ]}}\rightarrow \ \ \tau_{\mbox{\young[2,2][4][ , , , ]}}
\rightarrow \ \ \tau_{\mbox{\young[1,1,1][4][ , , , ]}}\rightarrow \ \ \tau_{\phi},\\
\ \ \dots \dots &:  \ \  \ \ \dots,
\end{cases}
\end{equation}
where $\tau_{\phi}=1$.

$(1,2)$-BTH  has  two sets of $\tau$-functions, and the Young diagram for each
$\tau$-function has one jump between adjacent rows:
\begin{equation}
\begin{cases}
(1,2)_{3\times 3}&:\ \ \tau_{\mbox{\young[4][4]}}\rightarrow \tau_{\mbox{\young[3,2][4]}} \rightarrow  \tau_{\mbox{\young[2,1][4]}},\\
&\ \ \ \tau_{\mbox{\young[5][4]}}\rightarrow   \tau_{\mbox{\young[4,3][4]}}\rightarrow\ \ \ \ \tau_{\mbox{\young[3,2,1][4]}},
\\
(1,2)_{4\times 4}&:
\ \ \tau_{\mbox{\young[6][4]}}\rightarrow\ \ \tau_{\mbox{\young[5,4][4]}}\rightarrow \ \ \tau_{\mbox{\young[4,3,2][4]}}\rightarrow \tau_{\mbox{\young[3,2,1
][4]}},\\
&\ \ \ \tau_{\mbox{\young[7][4]}}\rightarrow \ \ \tau_{\mbox{\young[6,5][4]}}\rightarrow \ \ \tau_{\mbox{\young[5,4,3][4]}}\rightarrow \ \ \tau_{\mbox{\young[4,3,2,1][4]}},\\
 \dots \dots &:  \ \  \ \ \dots.
\end{cases}
\end{equation}
Similarly $(1,3)$-BTH    has three sets of tau functions, and the Young diagram has two jumps between
adjacent rows.
\begin{equation*}
\begin{cases}
(1,3)_{4\times 4}&:\ \ \tau_{\mbox{\young[9][4]}}\rightarrow \ \ \tau_{\mbox{\young[8,6][4]}}\rightarrow\ \ \tau_{\mbox{\young[7,5,3][4]}}\rightarrow\ \ \tau_{\mbox{\young[6,4,2][4]}},\\
&\  \ \tau_{\mbox{\young[10][4]}}\rightarrow \ \ \tau_{\mbox{\young[9,7][4]}}\rightarrow\ \ \tau_{\mbox{\young[8,6,4][4]}}\rightarrow\ \ \tau_{\mbox{\young[7,5,3,1][4]}},\\
&\tau_{\mbox{\young[11][4]}}\rightarrow\ \ \tau_{\mbox{\young[10,8][4]}}\rightarrow\ \ \tau_{\mbox{\young[9,7,5][4]}}\rightarrow\ \ \tau_{\mbox{\young[8,6,4,2][4]}},\\
 \dots \dots &:  \ \  \ \ \dots.
\end{cases}
\end{equation*}
\end{Example}

\sectionnew{Conclusions and discussions}\label{sec:Conclusion}
 We proved the equivalence between $(N,M)$-BTH and $(M,N)$-BTH,  derived the primary Hirota equations of the $(N,M)$-BTH, and found several explicit formulas about solutions
 for the BTH  using orthogonal polynomials in the matrix form. We also constructed some rational solutions of the BTH which are parameterized by the products of Schur polynomials corresponding to non-rectangular Young diagrams.
 It may be interesting to find their significance in terms of the representation theory, as in the case of the
 original Toda lattice where the rational solutions are given by the Schur polynomials of rectangular Young diagrams and they are the Virasoro singular vectors.
\\

{\bf {Acknowledgments:}}
  {\small This work was carried out under the guidance of Professor Yuji Kodama during my visit to Ohio State University. I would like to thank Professor Yuji Kodama for his guidance and many useful discussions. I would also
  like to thank Department of Mathematics at Ohio State University for providing me a generous support and
   making my visit so pleasant.  I also thank Professor Jingsong He(NBU, China) for useful discussions
  and his general support.}



\end{document}